\documentclass[aps,amssymb,ams,twocolumn]{revtex4}
\usepackage[latin1]{inputenc}
\usepackage{amsmath}
\usepackage{amsfonts}
\usepackage{amssymb}
\usepackage{graphicx}
\usepackage{color}
\usepackage{natbib}
\usepackage[toc]{appendix}

\def\dotP#1#2{\vec{#1}\cdot\vec{#2}}		

\def\tensor#1{{\mathbf{#1}}}					                           
\def\fttensor#1{{\mathbf{\tilde{#1}}}}

\def\fft#1{\frac{1}{\sqrt{N}}\sum_{\vec{R}}e^{i\dotP{k}{R}} \;\tensor{#1}(\vec{R})}
\def\ifft#1{\frac{1}{\sqrt{N}}\sum_{\vec{k}}e^{-i\dotP{k}{R}}\;\fttensor{#1}(\vec{k})}

\def\mean#1{\left<{}#1\right>}	       

\begin{document}

\title{Inferring elastic properties of an fcc crystal from displacement correlations:\\sub-space projection and statistical artifacts}
\author{A. Hasan$^{(1)}$, C. E.~Maloney$^{(1)}$}
\affiliation{$^{(1)}$ Dept. of Civil and Environmental Engineering, Carnegie Mellon University, Pittsburgh, PA, USA}

\begin{abstract}
We compute the effective dispersion and vibrational density of states (DOS) of two-dimensional sub-regions of three dimensional face centered cubic (FCC) crystals using both a direct projection-inversion technique and a Monte Carlo simulation based on a common underlying Hamiltonian.
We study both a (111) and (100) plane.
We show that for any given direction of wavevector,  both (111) and (100) show an anomalous  $\omega^2\sim q$ regime at low $q$ where $\omega^2$ is the energy associated with the given mode and $q$ is its wavenumber.
The $\omega^2\sim q$ scaling should be expected to give rise to an anomalous DOS, $D_\omega$, at low $\omega$: $D_\omega \sim \omega^3$ rather than the conventional Debye result: $D_\omega\sim \omega^2$.
The DOS for (100) looks to be consistent with $D_\omega \sim \omega^3$, while (111) shows something closer to the conventional Debye result at the smallest frequencies.
In addition to the direct projection-inversion calculation, we perform Monte Carlo simulations to study the effects of finite sampling statistics.
We show that \emph{finite sampling} artifacts act as an effective disorder and bias $D_\omega$, giving a behavior closer to $D_\omega \sim \omega^2$ than $D_\omega \sim \omega^3$.
These results should have an important impact on the interpretation of recent studies of colloidal solids where the two-point displacement correlations can be obtained directly in real-space via microscopy.
\end{abstract}

\maketitle

\section{Introduction}
\label{section: introduction}

Colloidal suspensions of spherical particles exhibit similar behavior to conventional condensed matter systems~\cite{vanMegen:1986Nature}.
Direct observation of colloidal particle trajectories using optical microscopy has been employed to study condensed matter phenomena like: crystal nucleation~\cite{Weitz:2001science}, impurity frustrated crystallization~\citep{Volkert:2005science}, glass transitions~\citep{Weeks:2000science}, melting~\cite{Islam:2005science} etc.
When quenched into a solid-like state, particles fluctuate about their equilibrium  positions~\cite{Brito:2010SoftMat, Kaya:2010science, ChenLiuYodh:2010prl, Ghosh:2010prl}. 
At long wavelength these fluctuations should be governed by some effective elasticity and one should be able to extract the corresponding moduli from observed displacement fluctuations.

There are several ways to extract moduli from experimentally observed displacement fluctuations.  
Zahn {\it et. al.}~\cite{Zahn:2003prl} obtained bulk elastic constants by studying strain fluctuations in sub-regions of a 2D hexagonal colloidal crystal of increasing size and extrapolated to the infinite system size limit.
\citet{Keim:PRL2004} performed a planewave decomposition of particle displacements to compute the dispersion and used it to extract long wavelength elastic constants.
However these methods only work in spatially homogeneous systems.
Much like light scattering methods~\cite{ChengChaikin:2000prl, Hurd:pra1982},  they average out any spatial disorder.
In disordered systems, like structural glasses or geometrically ordered systems with heterogeneous interactions~\cite{Ganter:1998prl, Taraskin:2001prl}, the spatial heterogeneity and its impact on the fluctuation spectrum is of central interest, so other methods must be employed.

To study disordered systems, a third method has been employed by several groups~\cite{Kaya:2010science, ChenLiuYodh:2010prl, Ghosh:2010prl, Henkes:2012softMat}.
This approach exploits a connection between the linear elastic response function and displacement covariance matrix for a system in thermal equilibrium: $G_{i\alpha{}j\beta}=\mean{u_{i\alpha}u_{j\beta}}/k_{b}T$~\cite{Mazenko} where $k_b$ is Boltzmann's constant and $T$ is the temperature.
Here $u_{i\alpha}$ is the displacement of the $i$-th particle in the $\alpha$-th direction and the angle brackets mean thermal average.
In glassy colloids it has been used to study: the DOS~\cite{Ghosh:2010prl, Ghosh:2010softMat},  the  so-called Boson peak, an excess in the DOS when compared to the Debye theory,~\cite{ChenLiuYodh:2010prl} and the connection between structural relaxation and low frequency normal modes~\cite{ChenLiuYodh:2011prl, Ghosh:2011prl}.
For colloidal crystals, the method has been used to characterize the spectrum of normal modes and DOS~\cite{Kaya:2010science, Kaya:2011pre}.
However this wealth of information comes at a price and requires good statistical estimates for \emph{all} entries of $G_{i\alpha{}j\beta}$.
This raises a number of problems related to statistical convergence~\cite{Henkes:2012softMat} that can lead to qualitative misinterpretation of the data from finite samples, as we will show below.

There is another serious difficulty to which this method is susceptible when applied to a three-dimensional (3D) colloidal system.
To assemble $G_{i\alpha{}j\beta}$, one needs many uncorrelated snapshots of the system.
In a typical microscopy experiment, one has access to a single plane of the material and can not make simultaneous observations of particles above or below. 
However the observed in-plane fluctuations in the patch are mediated by degrees of freedom off the imaging plane that are not observable.

Thus one is faced with two problems. 
First, it is not obvious what effect the embedding medium has on the Green's function assembled from measurements made only in a lower dimensional patch.
This difficulty arises even in the case of a perfect, homogeneous crystal.
Second, it is not clear how finite statistics bias the spectrum, especially in the presence of the restricted window of observation. 

Recently Maggs and coworkers have made good progress on the first question~\cite{Maggs:2012epl, Maggs:2011epj-aniso, Maggs:2012softMat, Ghosh:2011physicaA}.
They have shown, most dramatically, that the expected energy of plane wave fluctuations in the restricted 2D patch is linear in wavenumber rather than quadratic as is the case for systems governed by a Laplace-like operator~\cite{AshcroftMermin}. 
However, their approach starts with a long wavelength cubic elasticity tensor and does not treat short wavelength fluctuations near the Brillouin Zone (BZ) boundary in a way that takes into account the structure of the underlying lattice. 
This makes it difficult to separate artifacts in the DOS near van Hove peaks into those arising from the observations being restricted to a 2D patch and those arising from finite sampling.
Furthermore, in the continuum elasticity approach, the anisotropic component of the elastic modulus tensor must be set by hand rather than emerging naturally from the underlying lattice structure and particle interactions. 
 
In this work, for a perfectly homogeneous face centered cubic (FCC) crystal with harmonic, nearest neighbor interactions, we first examine the dispersion relation and normal mode structure of the Green's function assembled from observing only in-plane fluctuations of a crystal patch.
Then we investigate the artifacts induced by finite sampling statistics.

In contrast to the approach of Maggs and coworkers, we work in a fully atomistic framework.
This allows us to produce realistic results for the entire Brilloiun zone and calculate the exact dispersion relation, frequency spectrum and normal modes for finite 2D patches cut from $(100)$ and $(111)$ FCC crystal planes.
Our dispersion results agree in the long wavelength limit with those in refs~\cite{Maggs:2012epl, Maggs:2011epj-aniso}.
However, in the rest of the Brillouin zone we find significant qualitative differences between the longitudinal and transverse branches of the dispersion.
Moreover, in the low frequency portion of the $[111]$ patch DOS, we see pronounced deviations from the scaling expected from the long wavelength dispersion relation.

At the same time we perform a Monte-Carlo simulation using \emph{precisely the same} Hamiltonian to make a direct, quantitative comparison between the true projected spectrum and those spectra inferred from finite statistical samples.
We find that finite statistics induces important artifacts in the DOS: it smears out and shifts the van Hove singularities to lower frequency parts of the spectrum in a manner similar to disorder~\cite{Taraskin:2001prl}.
This alters the inferred low frequency scaling behavior in a qualitative way.


\section{Elasticity of a two dimensional subsystem}
\label{section: elasticity of two diemsional subsystem}

\begin{figure}[b]    
\begin{center}
\includegraphics[width=.15\textwidth]{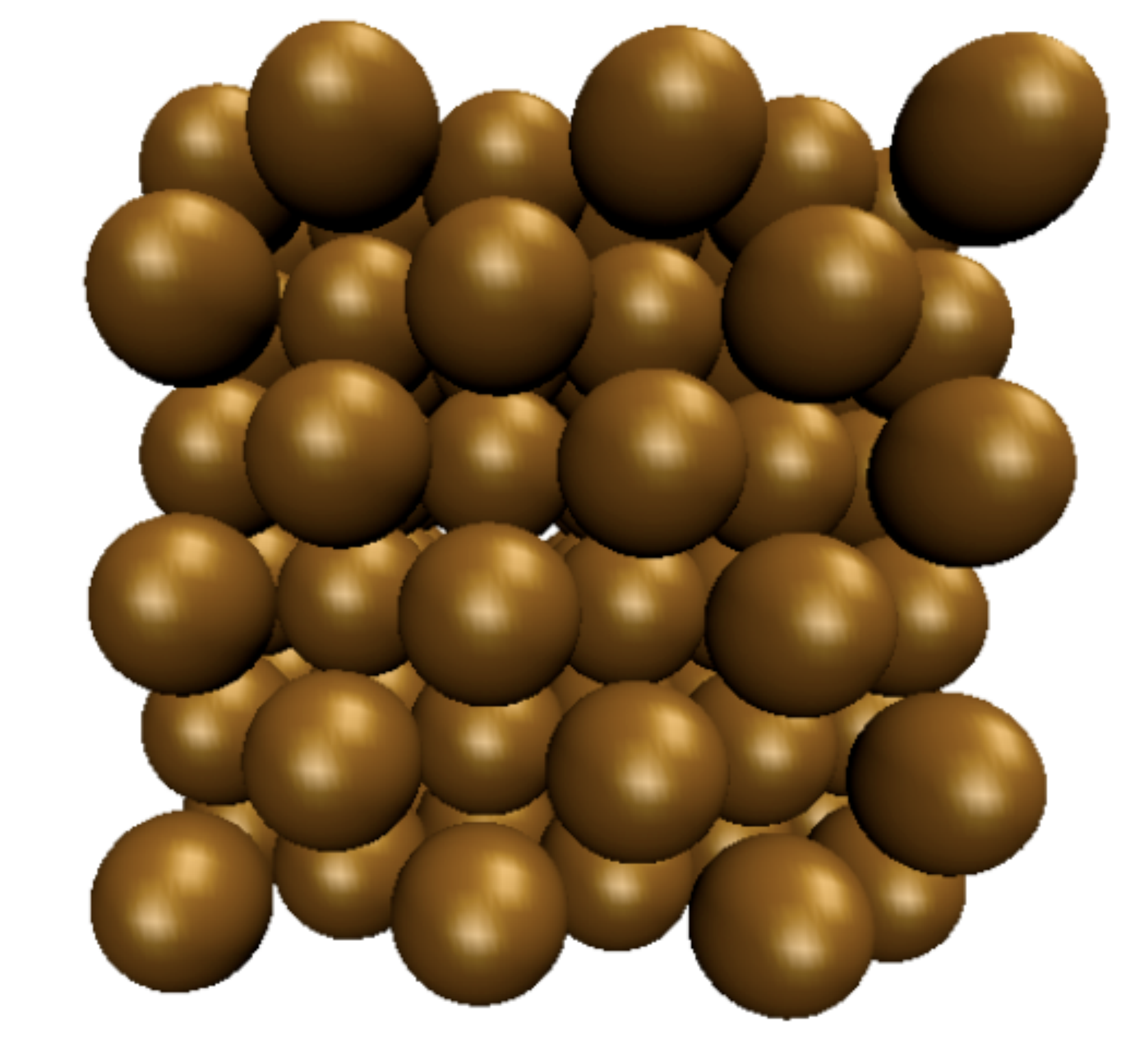} 
\includegraphics[width=.15\textwidth]{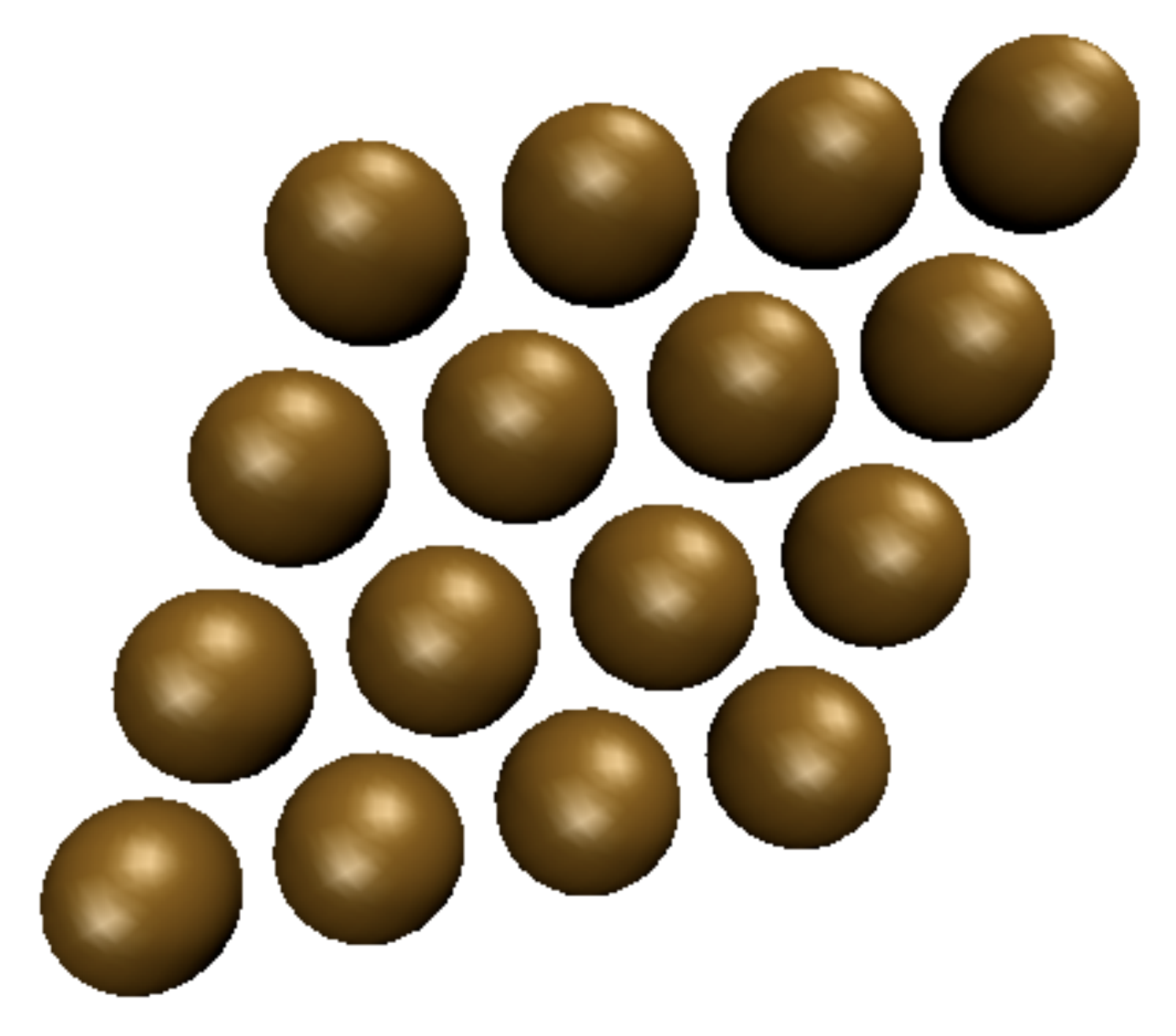} 
\includegraphics[width=.15\textwidth]{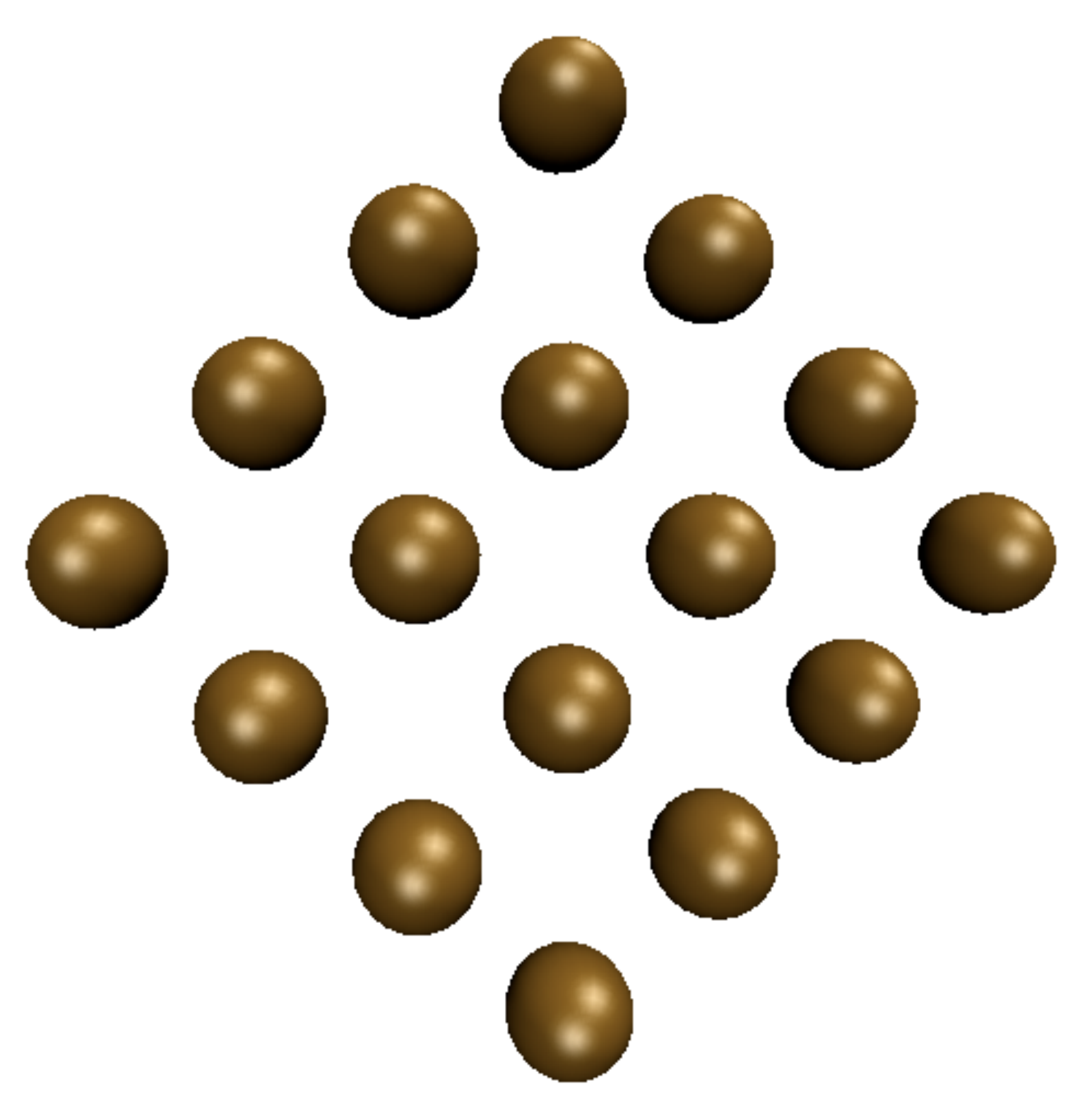}
\end{center}
\caption{
Left: FCC crystal with $C=3$ cubic unit cells along each edge.
Center: Rhombic $(111)$ FCC patch with $P=4$ atoms along each edge.
Right: Square $(100)$ FCC patch with $P=4$ atoms along each edge.
}
\label{fig: slicing pics}
\end{figure}

We analytically compute the elastic Green's function of a 2-dimensional patch of atoms embedded in homogeneous, face-centered-cubic (FCC) crystal with cubic periodic boundaries.     
The domain and range of the Green's function of a patch, denoted by $\mathcal{G}$, of $M$ atoms is a $\mathbb{R}^{2M}$ subspace of the $\mathbb{R}^{3N}$ configuration space of the FCC crystal (having $N$ atoms).
The embedding FCC crystal is governed by a harmonic Hamiltonian having only, pair-wise interactions, connecting nearest neighbors with unstressed springs. 
We denote the nearest neighbor spacing by $a$ and measure all lengths as units of $a$.
For this system the bond stiffness tensor is: $B_{i\alpha{}j\beta} = K\hat{r}_{ij\alpha}\hat{r}_{ij\beta}$~\cite{AshcroftMermin}. 
Here $K$ is the bond strength;  Latin indices $i,j$ label atoms, while Greek indices $\alpha{},\beta$ label Cartesian axes.
$\hat{r}_{ij}$ is the unit vector pointing from atom $i$ to one of its nearest neighbors, atom $j$. 
The non-zero entries of the Hessian matrix, $\mathbb{H}$, are given by: $H_{i\alpha{}j\beta} = B_{i\alpha{}j\beta}$ when $i\neq{}j$, and $H_{i\alpha{}j\beta} = -\sum_{j,j\neq{}i}B_{i\alpha{}j\beta}$ for $i=j$. 
            
To obain the Green's function, $\mathcal{G}$, of a $P\times{}P$ atom patch, we start by computing the full Green's function $\mathbb{G}=\mathbb{H}^{-1}$, of the embedding cubic, periodic crystal with $C$ 4-atom cubic unit cells along each edge. 
The computation of $\mathbb{G}$ nominally amounts to a computationally expensive, inversion of the operator $\mathbb{H}$. 
However, due to periodicity, $\mathbb{H}$ is a block-circulant matrix, and we use a Fourier space approach (detailed in appendix~\ref{section: Green's functions analytical expressions}) which permits inversion in linear computer time and space. 
In practice we have used this method to invert Hessians of roughly 8 million particle systems in a few tens of minutes of computer time on a single workstation.
Projecting $\mathbb{G}$ onto the observed degrees of freedom in the patch, we immediately obtain $\mathcal{G}$.

We next discuss various properties (dispersion, mode structure and DOS) of $\mathcal{G}$ for patches cut from the (111) and (100) planes of an FCC crystal.
The embedding crystal has $C$ cubic unit cell cells along each edge, while the patch has $P$ atoms along each edge of its boundary.
Figure~\ref{fig: slicing pics} has an illustration of our patches.
We take the $(111)$ patch to be a triangle lattice with rhombic boundaries and the $(100)$ patch to be a square lattice with square bounds.

\subsection{Pseudo-dispersion}
\label{subsection: pseudodispersion}

\begin{figure}[b]    
\begin{center}
\includegraphics[width=.49\textwidth]{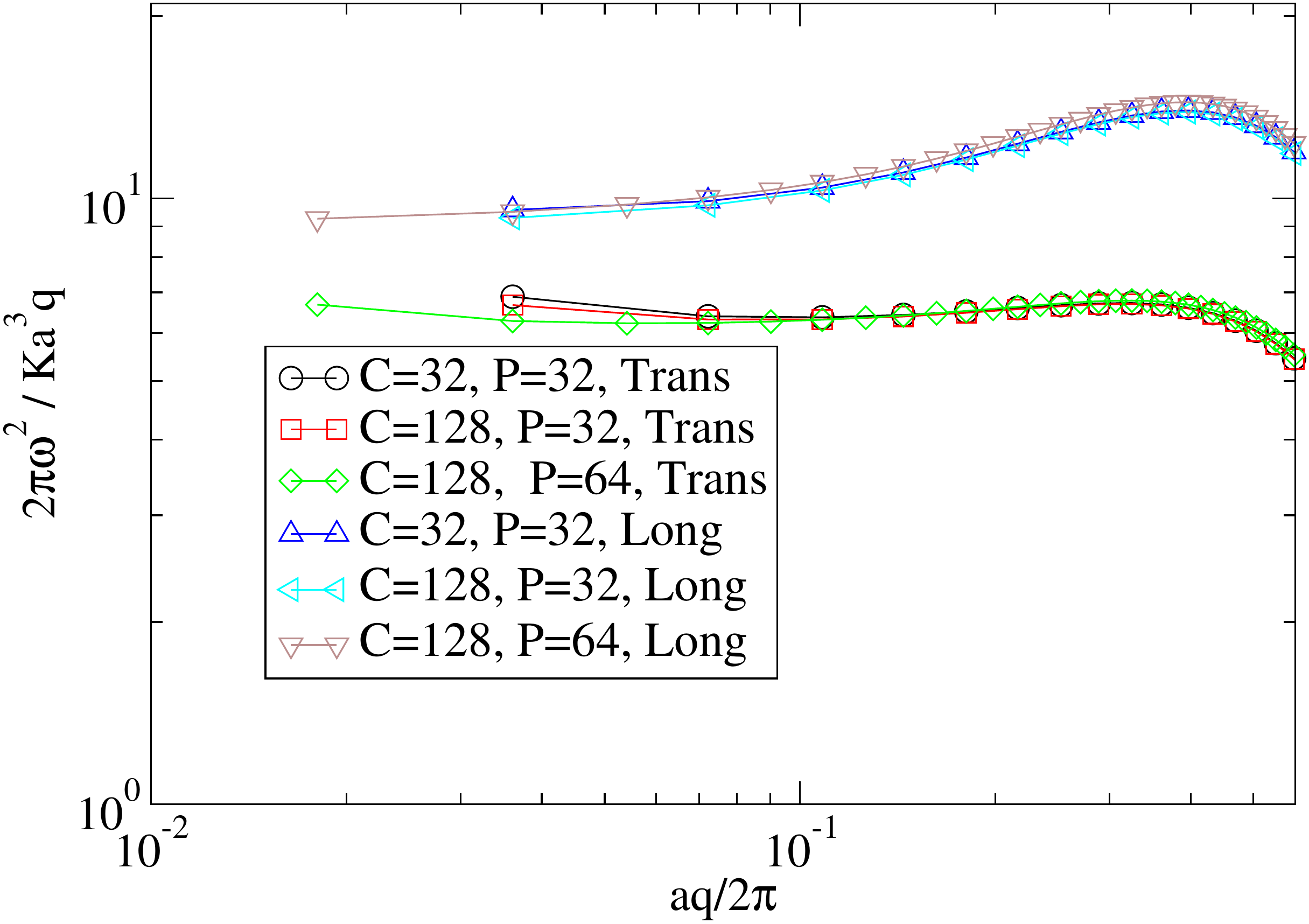}
\end{center}
\caption{
Computed longitudinal and transverse dispersion curves scaled by wavenumber for planewaves traveling along the next-nearest neighbor direction and lying in the first Brillouin zone for $(111)$ various patches.
}
\label{fig: dispersion [111] near-neighbor}
\end{figure}

\begin{figure}[th!]    
\begin{center}
\includegraphics[width=.49\textwidth]{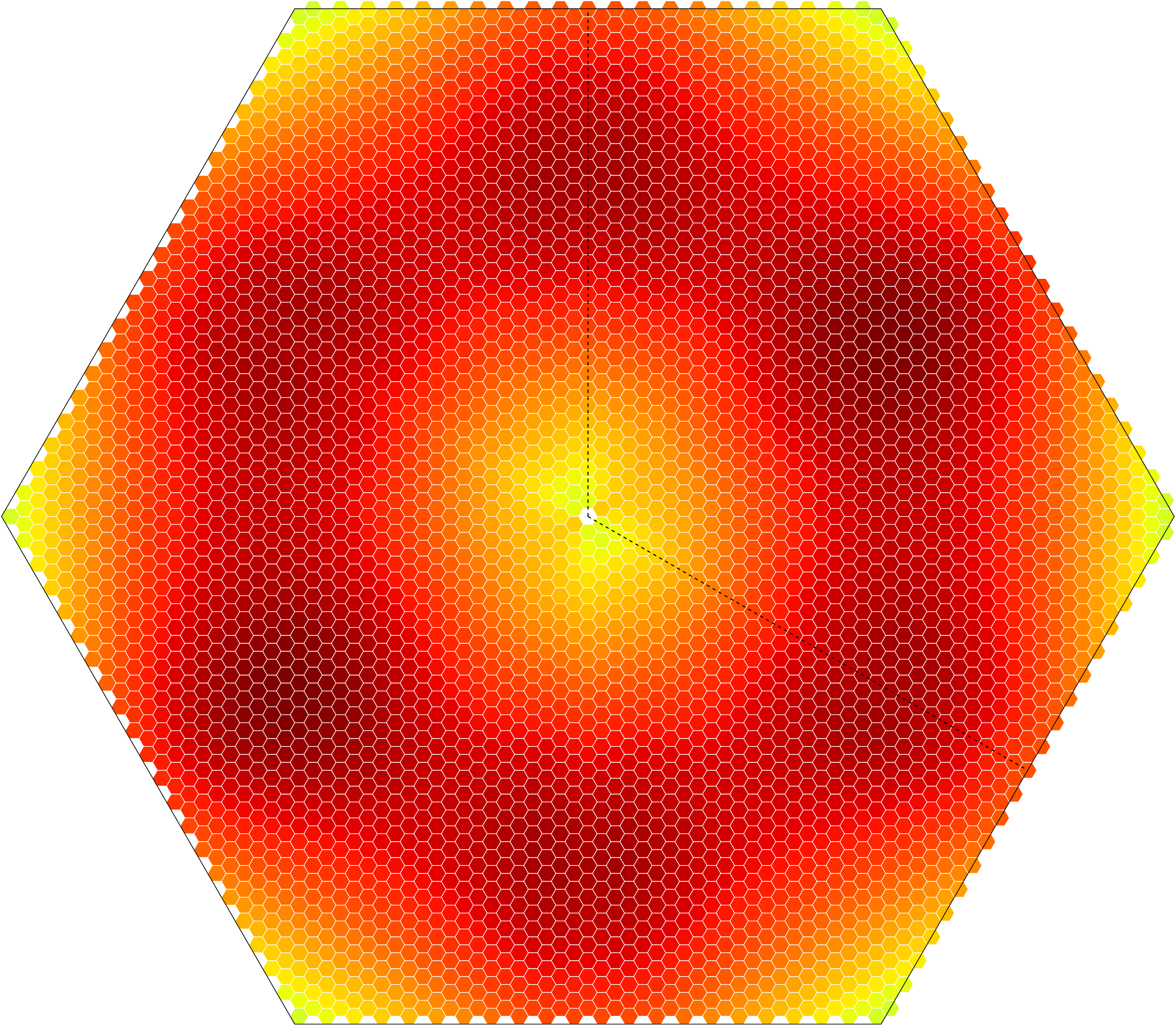} \\
\includegraphics[width=.49\textwidth]{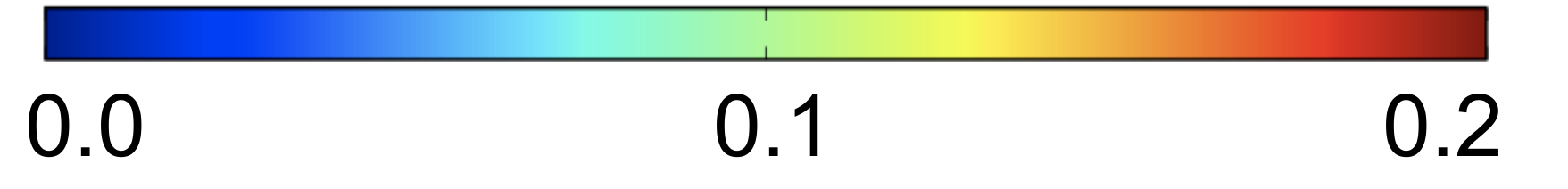} 
\includegraphics[width=.49\textwidth]{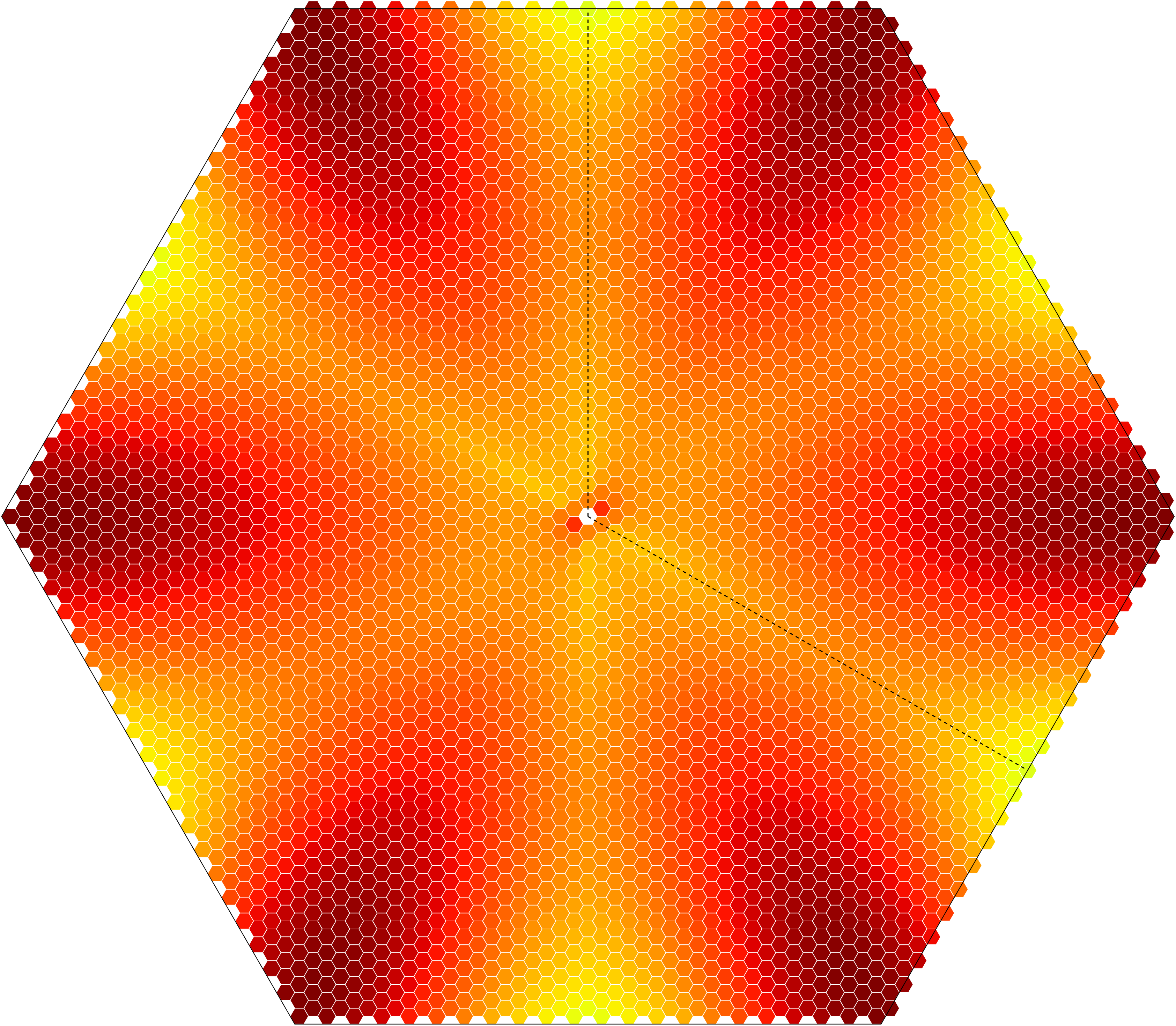} \\
\includegraphics[width=.49\textwidth]{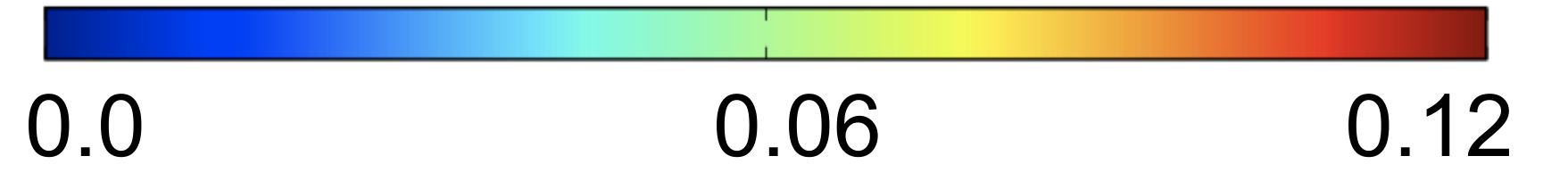} 
\end{center}
\caption{
Computed dispersion scaled by wavenumber, $\omega^{2}/Ka^{2}q$, for a $(111)$ patch ($P=64$, $C=64$) for waves in the first Brillouin zone.
Top: Longitudinal.  
Bottom: Transverse.
}
\label{fig: complete C32 [111] analytical dispersion}
\end{figure}

For each vector $\vec{q}$ that lies in the first BZ of the patch's reciprocal lattice, planewave excitations are given by:
$\psi_{i\alpha{}L(T)}(\vec{q}) = \hat{p}_{\alpha{}L(T)}e^{i\dotP{q}{r_{i}}}$; $\hat{p}_{\alpha}$ is the component of the polarization vector along the $\alpha$-axis.
Here the subscripts $L$ and $T$ label longitudinal and transverse polarizations respectively, while $\vec{r}_{i}$ labels the equilibrium position of atom $i$.
Because of boundary effects,  plane waves are not, strictly speaking, eigenmodes of $\mathbb{\mathcal{G}}$ even for high symmetry directions.
Nevertheless, we compute the longitudinal (transverse) \emph{pseudo-dispersion} $\omega^{2}(\vec{q})$ as:
\begin{equation}
\label{eqn: pseudodispersion}
1/\omega^{2}_{L(T)}(\vec{q})= \sum_{i\alpha{}j\beta}\psi_{i\alpha{}L(T)}\mathcal{G}_{i\alpha{}j\beta}\psi_{j\beta{}L(T)}
\end{equation} 
In all subsequent results, we report energies in units of $Ka^2$.

\begin{figure}[t]    
\begin{center}
\includegraphics[width=.49\textwidth]{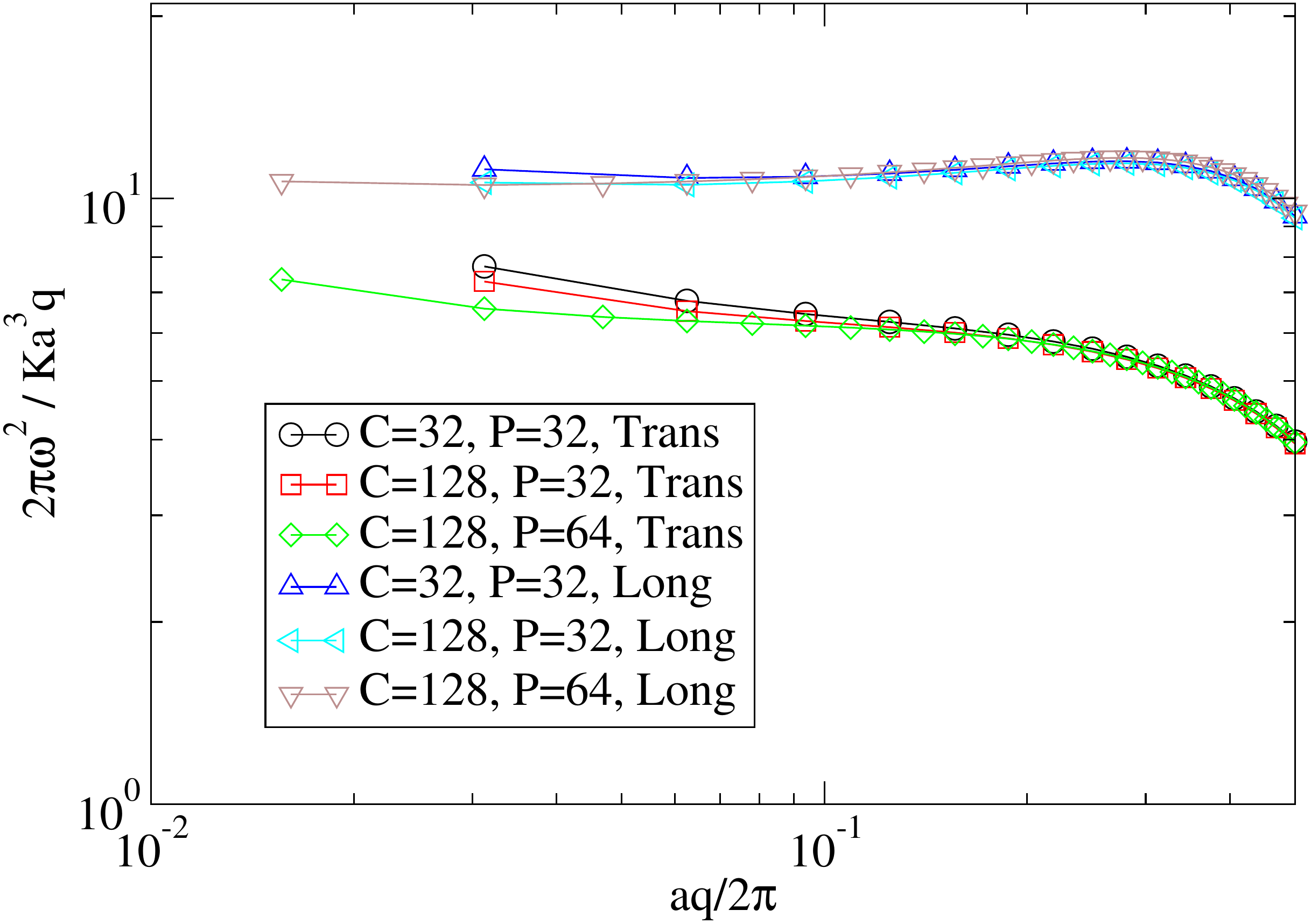} 
\includegraphics[width=.49\textwidth]{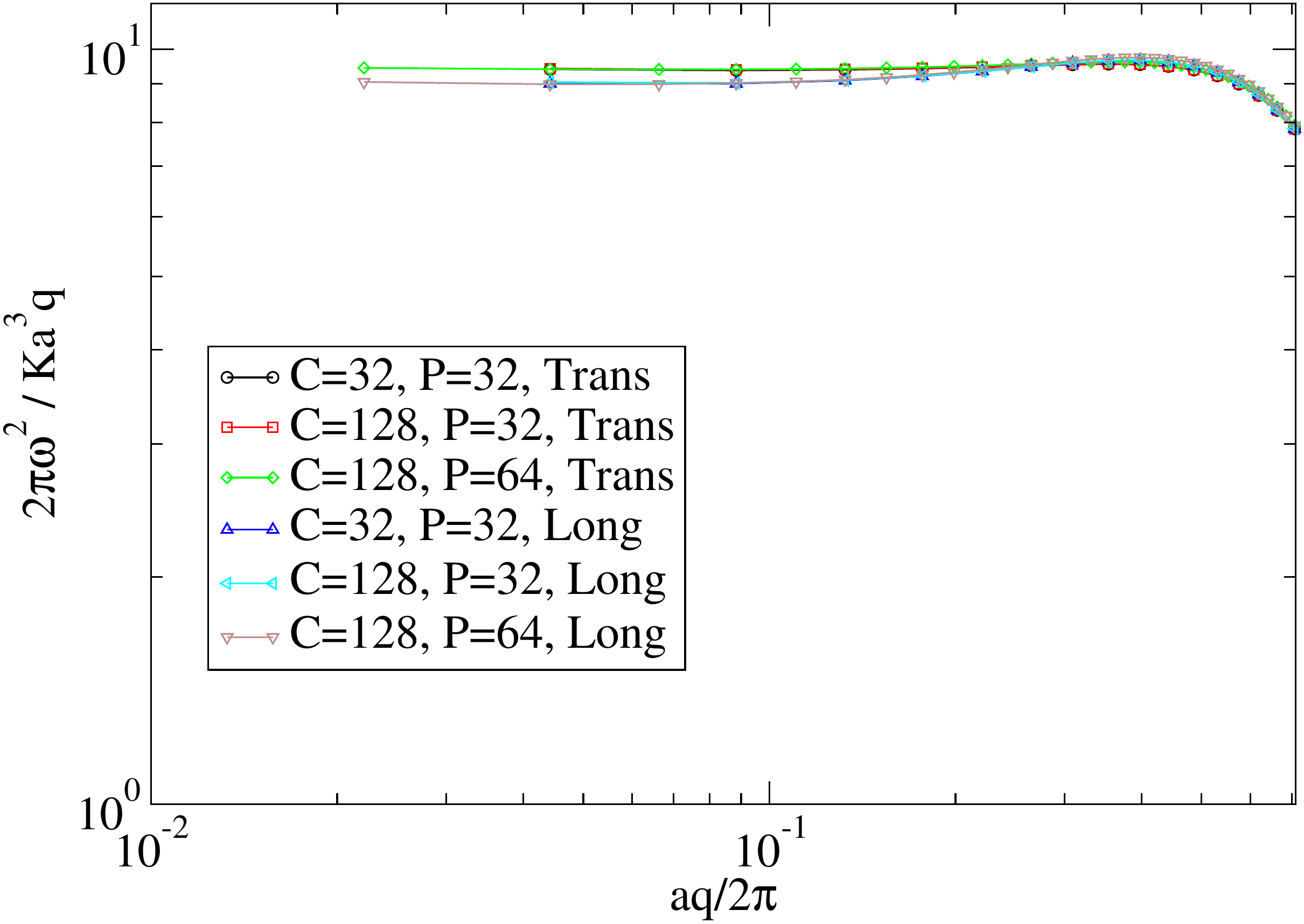} 
\end{center}
\caption{
Computed longitudinal and transverse dispersion curves scaled by wavenumber for planewaves lying in the first Brillouin zone of various $(100)$ patches.
Top: For planewaves travelling in the nearest neighbor direction ($[010]$ direction).
Bottom: For planewaves travelling in the next-nearest neighbor direction ($[011]$ direction).
}
\label{fig: dispersion [100] near-neighbor}
\end{figure}

\begin{figure}[th!]    
\begin{center}
\includegraphics[width=.47\textwidth]{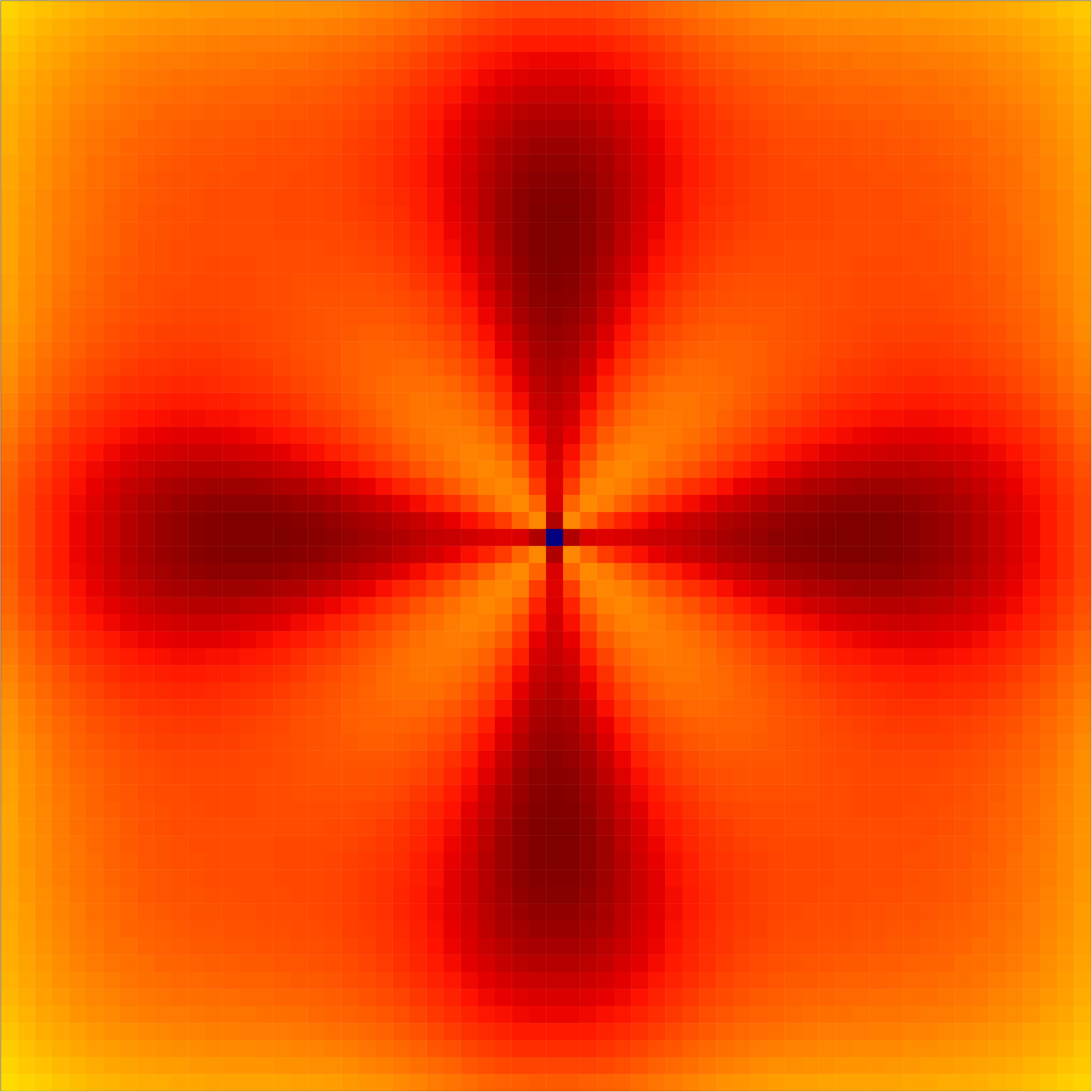} 
\includegraphics[width=.49\textwidth]{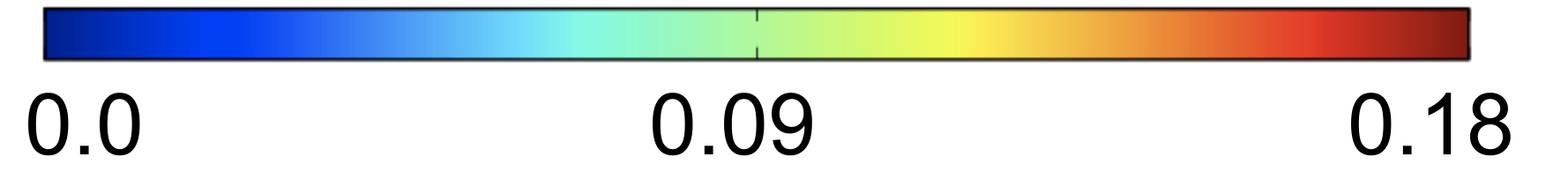}
\includegraphics[width=.47\textwidth]{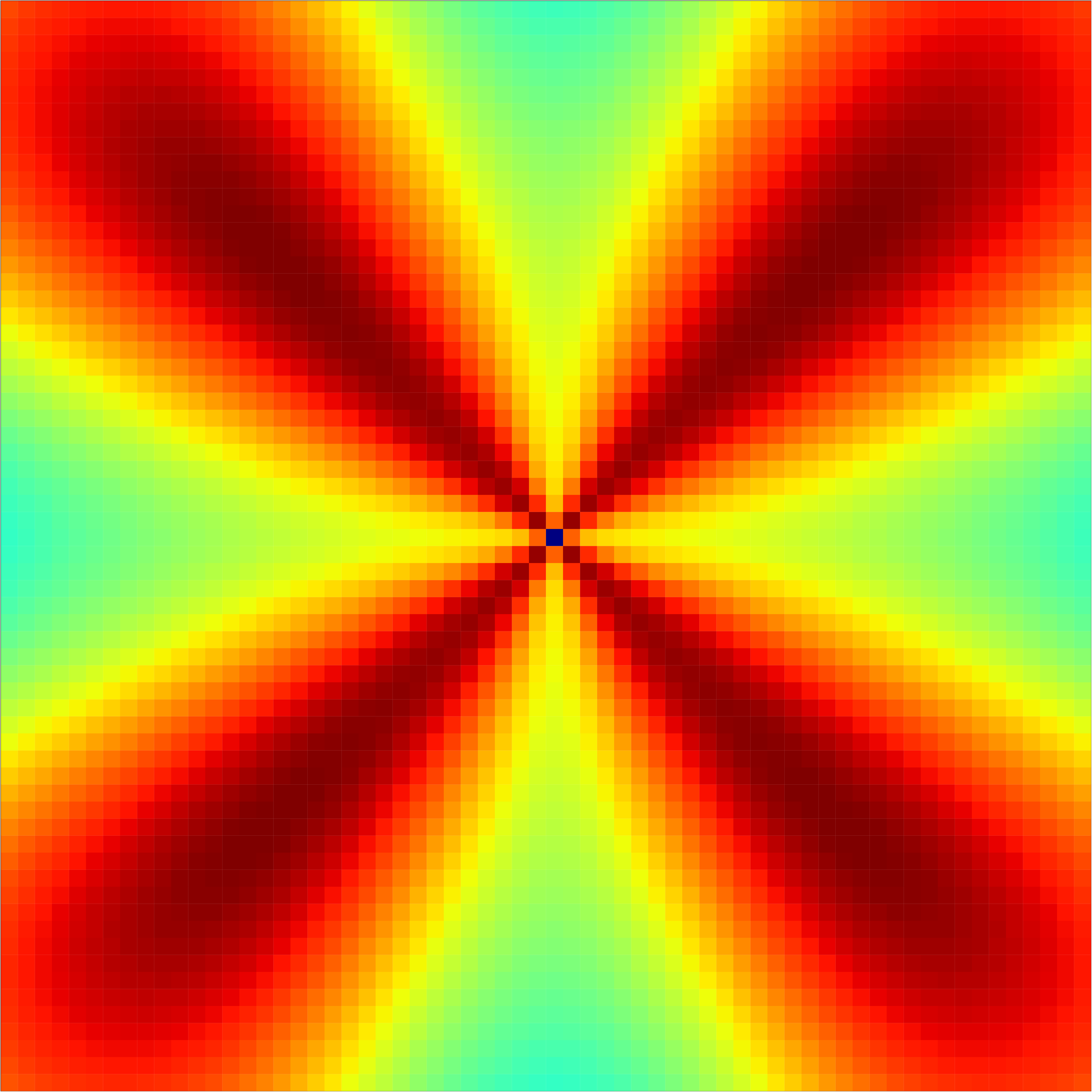} 
\includegraphics[width=.49\textwidth]{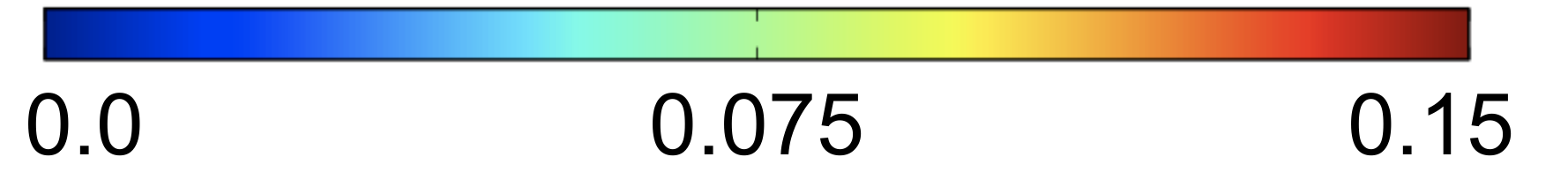}
\end{center}
\caption{
Computed dispersion scaled by wavenumber: $\omega^{2}/Ka^{2}q$, for a $(100)$ patch ($P=64$, $C=64$) for waves in the first Brillouin zone.
Top: Longitudinal.  
Bottom: Transverse.
}
\label{fig: complete C32 [100] analytical dispersion}
\end{figure}

An approximation for scalar elasticity governed by the Laplace operator in an infinite continuum~\cite{Ghosh:2011physicaA} as well as later work for continuum elasticity governed by a  long wavelength cubic elasticity tensor~\cite{Maggs:2012epl, Maggs:2011epj-aniso} predict the anomalous \emph{non-linear} dispersion of $\omega^{2}\sim{}q$ (at low $q$) for the patch Green's function $\mathcal{G}$.
Our approach differs from this previous work in that the underlying lattice and interactions dictate the symmetry of the long wavelength elastic modulus tensor and determine the behavior of the system at the BZ boundary -- in particular the magnitude and location of the van Hove peaks -- in a realistic way for an FCC crystal. 
In figure~\ref{fig: dispersion [111] near-neighbor} we show  $\omega^2/q$, for various patches cut from the $(111)$ plane and for longitudinal and transverse planewaves traveling along an edge of the BZ.
Note that a patch with $P=32$ atoms along an edge cut from a crystal of size $C=32$ is roughly half the size of the largest $(111)$ plane that fits in the embedding FCC crystal.
Figure~\ref{fig: dispersion [111] near-neighbor} shows that patches of size $P=32$ are already quite insensitive to the size of the full embedding crystal once the embedding crystal is just slightly larger than the patch itself.
We can also see that the ratio of the longitudinal to transverse branch is \emph{not} $3$ as it would be for a true 2D triangular lattice with central force interactions.

Figure~\ref{fig: complete C32 [111] analytical dispersion} shows the complete dispersion scaled by the wavenumber for wavevectors $\vec{q}$ that lie in the first BZ of a $(111)$ patch ($P=64$, $C=64$).
We see that at low $q$, there is little variation in the scaled dispersion along any particular radial direction in the BZ. 
The angular dependence of the dispersion has hexagonal symmetry, as it must.
Moreover at low wavenumbers the dispersion is isotropic - particularly in the longitudinal case.
A similar calculation indicates that the dispersion of a 2D triangular lattice has a similar angular dependence as that of the $(111)$ FCC patch.
Note that a slight breaking of hexagonal symmetry is apparent in figure~\ref{fig: complete C32 [111] analytical dispersion} near the center of the BZ.
This is an artifact of cutting a rhombic $(111)$ patch from a cubic FCC crystal: the edges of the rhombus are not crystallographically equivalent to its shorter diagonal.

The dispersion for square $(100)$ patches essentially follows the same non-linear scaling $\omega^{2}\sim{}q$  for low $q$, but with significant differences compared to $(111)$. 
Figure~\ref{fig: dispersion [100] near-neighbor} shows the dispersion relation for various $(100)$ patches and for planewaves traveling along the $[010]$ (top) and $[011]$ (bottom) directions.
We see that longitudinal and transverse planewaves traveling along a diagonal of the square ($[011]$ direction) are nearly degenerate (as in the Hard-Sphere Monte Carlo simulations of reference~\cite{Maggs:2011epj-aniso}), while the planewaves along the the $[010]$ direction are not.
This suggests that the $(100)$ patch dispersion is strongly anisotropic.
Figure~\ref{fig: complete C32 [100] analytical dispersion} shows the dispersion scaled by the wavenumber for wavevectors $\vec{q}$ that lie in the first BZ  of a $(100)$ patch ($P=64$, $C=64$).
The scaled dispersion shows strong anisotropy that is particularly pronounced in the transverse branch.
For the transverse case, $\omega^2/q$ also shows \emph{radial} variations at low $q$ along some, but not all, directions indicating that the low-$q$ scaling is \emph{not} as robust in every direction, with the $[010]$ direction giving particularly poor scaling at small $q$.

\subsection{Density of States}
\label{subsection: Density of States}
The Debye prediction for the low frequency scaling of$D_{\omega}=dN/d\omega$, the number of modes \emph{per unit $\omega$}, is: $D_{\omega}\sim\omega^{d-1}$ in $d$ dimensions.
However, this scaling assumes that $\omega^2\sim q^2$ at low frequency.
On one hand, it has been pointed out that the anomalous low frequency dispersion of $\mathcal{G}$, $\omega^{2}\sim{}q$ for low $q$, will change the expected scaling to $D_{\omega}\sim\omega^{3}$~\cite{Ghosh:2011physicaA}.
On the other hand, experimental data~\cite{Kaya:2010science} and some hard-sphere Monte Carlo simulation results~\cite{Maggs:2011epj-aniso} for $(111)$ seem to suggest that $D_{\omega}\sim\omega^{2}$ at finite but low $\omega$, in accord with the standard Debye prediction for a 3D material.
We now compute the DOS, of finite rhombic $(111)$ and square $(100)$ patches cut from the full FCC crystal, paying special attention to the low frequency scaling behavior.

\begin{figure}[b]    
\begin{center}
\includegraphics[width=.49\textwidth]{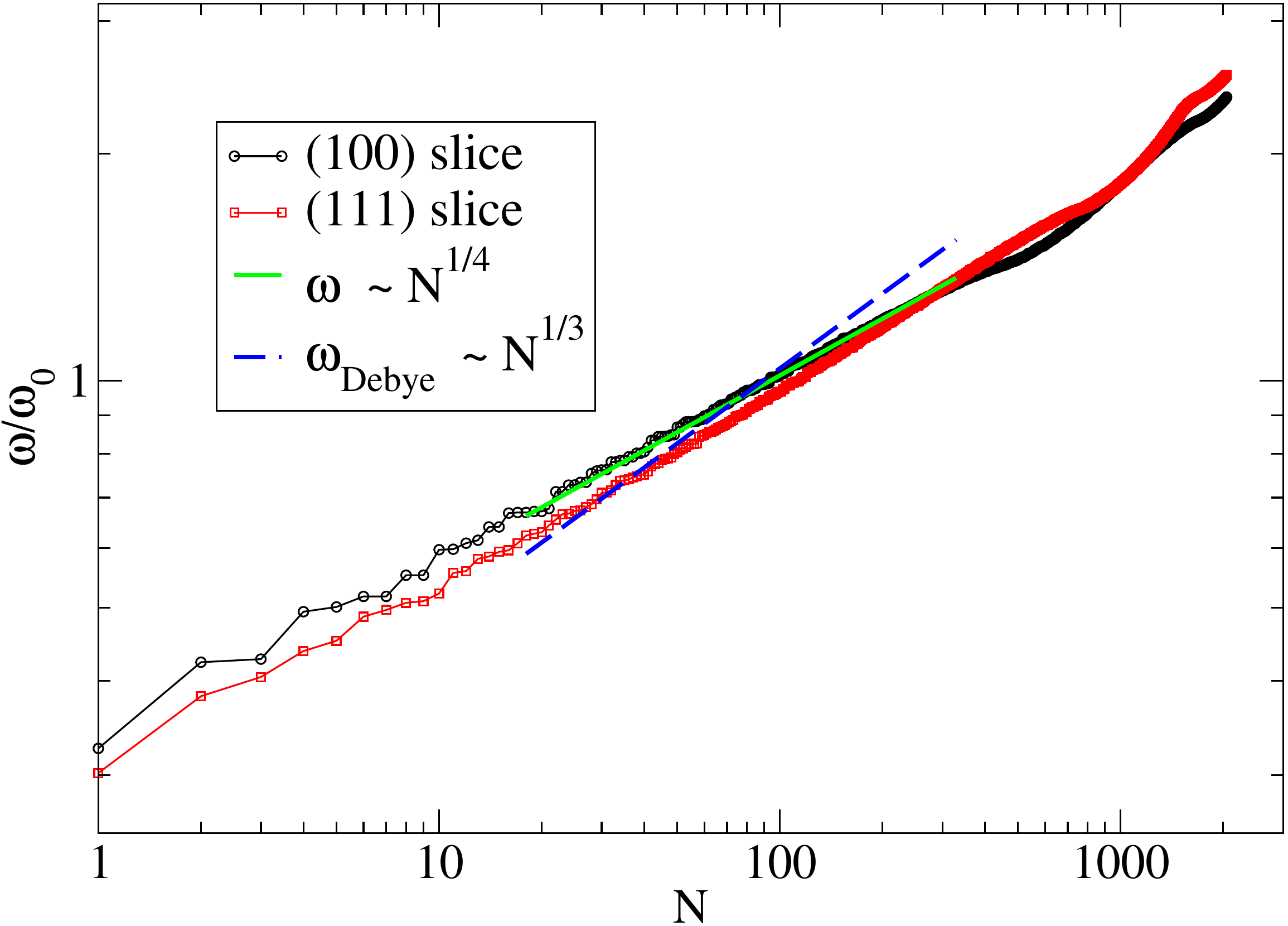}
\end{center}
\caption{
Integrated DOS $N(\omega)=\int^{\omega}D_{\omega}d\omega$. Frequencies, $\omega$, as a function of index, $N$, for $(100)$ and $(111)$ FCC patches ($P=32$, $C=32$).
}
\label{fig: integrated DOS}
\end{figure}

\begin{figure}[t]    
\begin{center}
\includegraphics[width=.49\textwidth]{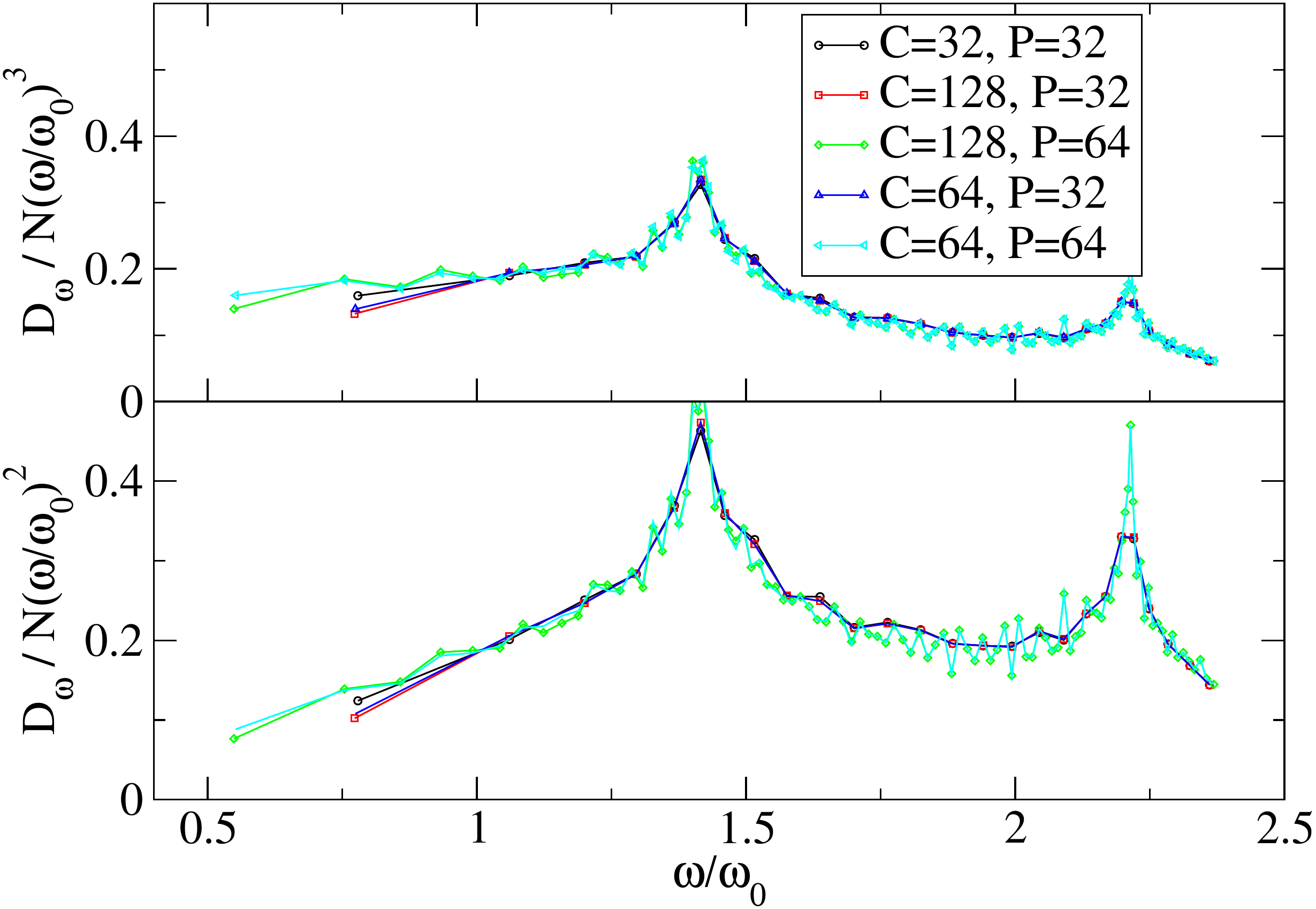}
\includegraphics[width=.49\textwidth]{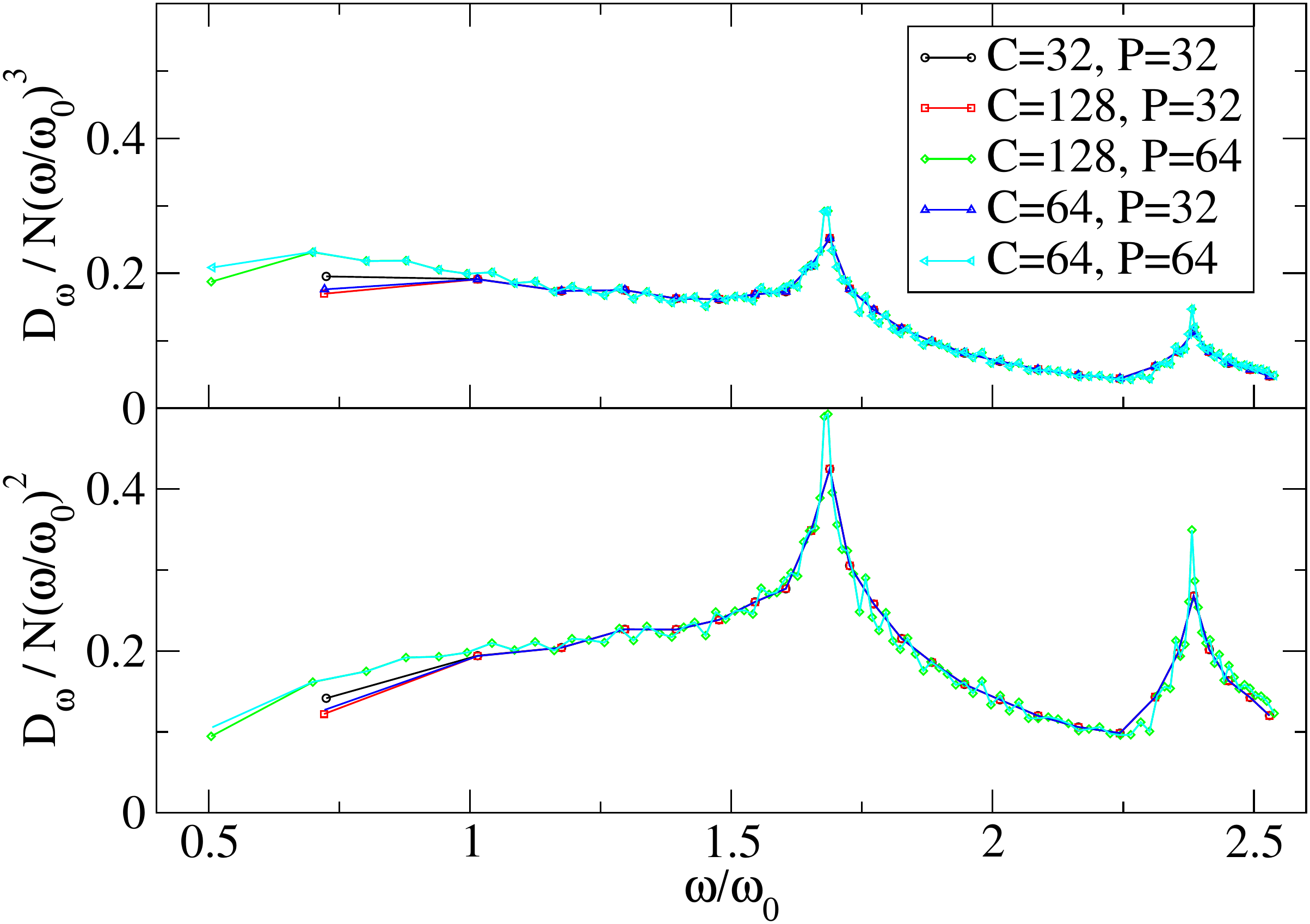}
\end{center}
\caption{
Computed DOS, $D_\omega$, normalized by the total number of normal modes, $N$, for various FCC crystal patches.
Top: Patch from $(100)$  crystal plane.
Bottom: Patch from $(111)$ crystal plane.
Upper panel: $D_\omega{}/N$ scaled by the non-Debye scaling $\omega^3$. 
Lower panel: $D_\omega{}/N$ scaled by the Debye scaling $\omega^2$.
}
\label{fig: DOS computed}
\end{figure}

We first compute the set of eigenvalues of $\mathcal{G}$, $\lambda_{p}$ where $p$ indexes the mode.  
The $p$-th.~frequency is then given by $\omega_{p}=1/\sqrt{m\lambda_{p}}$ where $m$ is particle mass.
We report frequency in units of $\omega_{0}=\sqrt{K/m}$.
We first consider the $\emph{integrated}$ DOS: $N(\omega)=\int_0^{\omega}D_{\omega}d\omega$.
Figure~\ref{fig: integrated DOS} shows $N(\omega)$ for $(111)$ and $(100)$  patches with $P=32$ atoms along each edge embedded in a FCC crystal with $C=32$ cubic unit cells along each edge. 
In addtion, we have overplotted the conventional Debye scaling $\omega\sim{}N^{1/3}$ and the expected modified Debye scaling: $\omega\sim{}N^{1/4}$, for an intermediate.
For both the patches, we see that the conventional Debye scaling does worse than the new modified-Debye prediction.
However, $N(\omega)$ of the $(100)$ patch shows almost perfect agreement with the expected non-Debye scaling from $N=10$ to about $N=200$, whereas the $(111)$ patch deviates by as much as $20$\% from the expected non-Debye scaling even in this low-$\omega$ regime.
We verify that our results are qualitatively unchanged if patch size $P$ is kept constant while the embedding FCC crystal's size $C$ is increased.
That is, the size of the actual system has minimal impact on the results, while the size of the observation window must induce some finite size effects.
We suspect that increasing the patch size would extend the  modified-Debye scaling regime to lower frequencies, but that one would always observe departures by at least $20\%$ for $\omega > 0.6\omega_0$.

We next make a histogram with the computed frequencies to directly estimate $D_{\omega}$ for various FCC patches and check scaling behavior at low $\omega$. 
Figure~\ref{fig: DOS computed} shows the DOS for $(100)$ (top) and $(111)$ (bottom) patches scaled by $\omega^{3}$ and $\omega^{2}$.
In each case $D_{\omega}$ is normalized by the total number of normal modes so that area under the graph is unity.
Our histogramming algorithm uses bins of variable width in $\omega$-space so that the number of points lying in each bin is the same.
For consistency we use $80$ points per bin in all our histogram plots. 
We have checked that the location of the van Hove peaks and low frequency scaling behavior of $D_{\omega}$ is independent of these choices.
We see from the plots of figure~\ref{fig: DOS computed} that for $C>P$ the embedding crystal is large enough such that $C$ does not directly impact the results.
The $(100)$ patch DOS, when scaled by the expected non-Debye behavior, $\omega^{3}$, shows a plateau for low $\omega$, while it lacks a plateau when scaled by the conventional Debye  scaling, $\omega^{2}$; in accord with the integrated DOS in figure~\ref{fig: integrated DOS}. 

In contrast to the $(100)$ patch, the $(111)$ behavior (figure~\ref{fig: DOS computed} bottom) is less clear.
For the largest patches studied, $P=64$, there is a clear bump in the plateau in  $D_\omega/\omega^3$ at the lowest frequencies.
This bump gives the conventionally scaled version,  $D_\omega/\omega^2$, something close to an apparent plateau at low frequency; something that was clearly not observed in the $(100)$ patch.
An experimenter, using the statistical approach described later, looking at a patch of this size would be hard pressed to decide whether the low frequency regime scaled like the expected $\omega^3$ or the conventional $\omega^2$.

In parting, we note that a similar plateau in $D_\omega /omega^2$ for a $(111)$ patch was observed in reference~\cite{Maggs:2011epj-aniso}.
However, the DOS in that case was computed from a displacement correlation analysis in a hard sphere simulation.
As we will show below, statistical artifacts themselves can act to give an even more robust plateau in $D_\omega /omega^2$ at intermediate frequencies, so it becomes difficult to say whether in the case of reference~\cite{Maggs:2011epj-aniso}, the apparent plateau was the result of projection artifacts or statistical artifacts.

\subsection{Mode Structure}
\label{subsection: mode structure}

\newcommand{\figwidth}{0.26}
\begin{figure*}[ph!]    
\begin{center}
\includegraphics[width=\figwidth\textwidth]{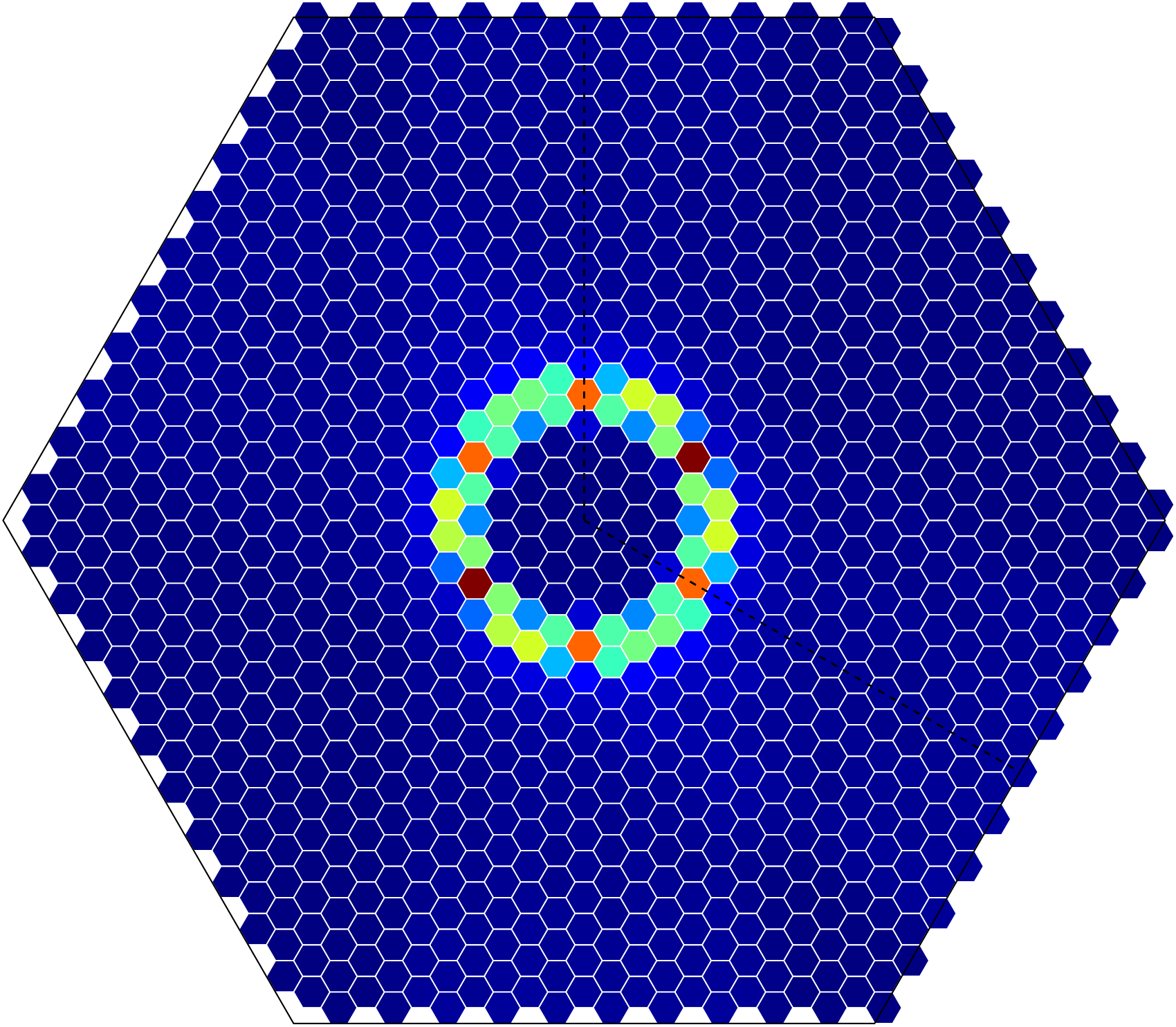} 
\includegraphics[width=\figwidth\textwidth]{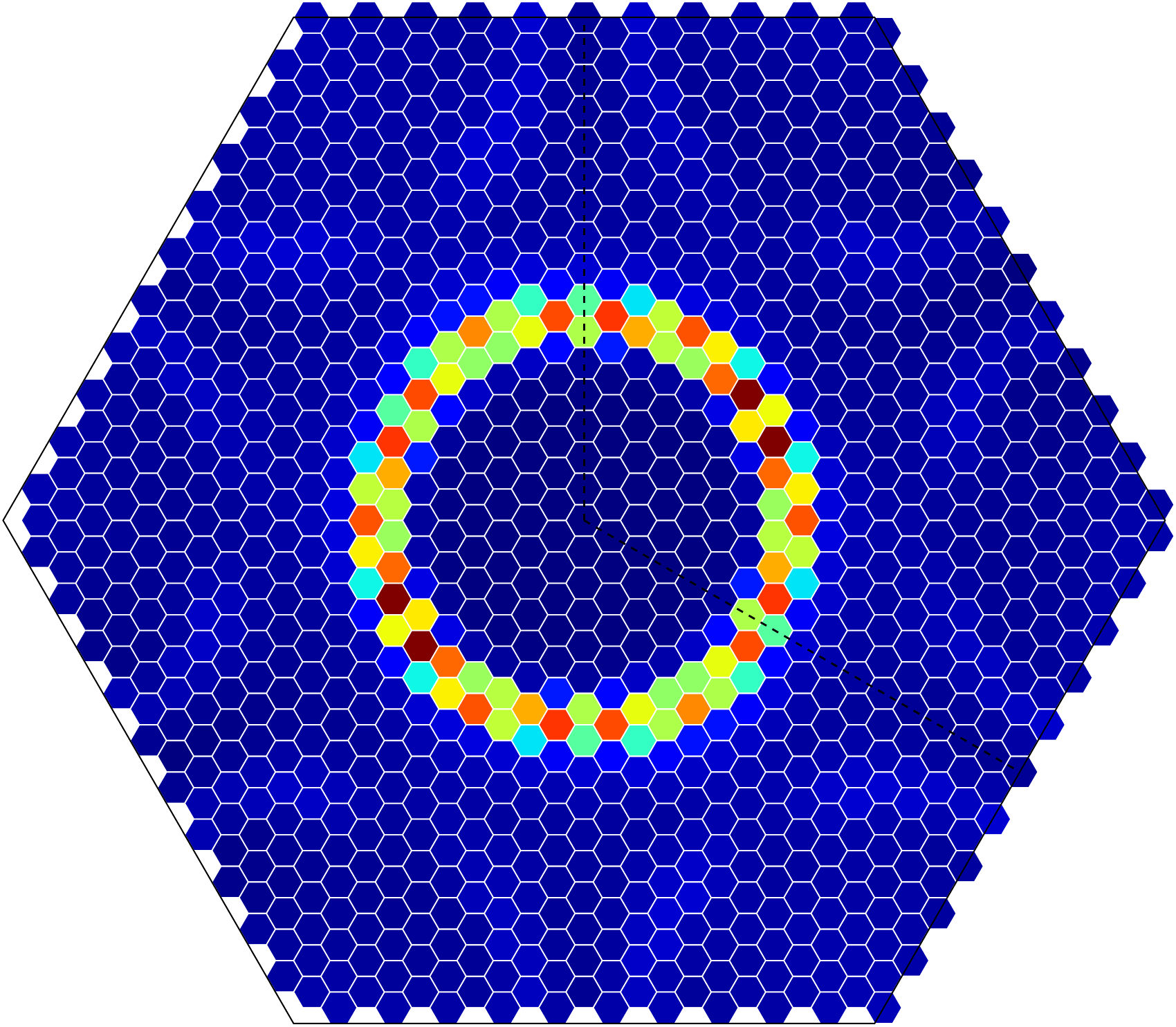} 
\includegraphics[width=\figwidth\textwidth]{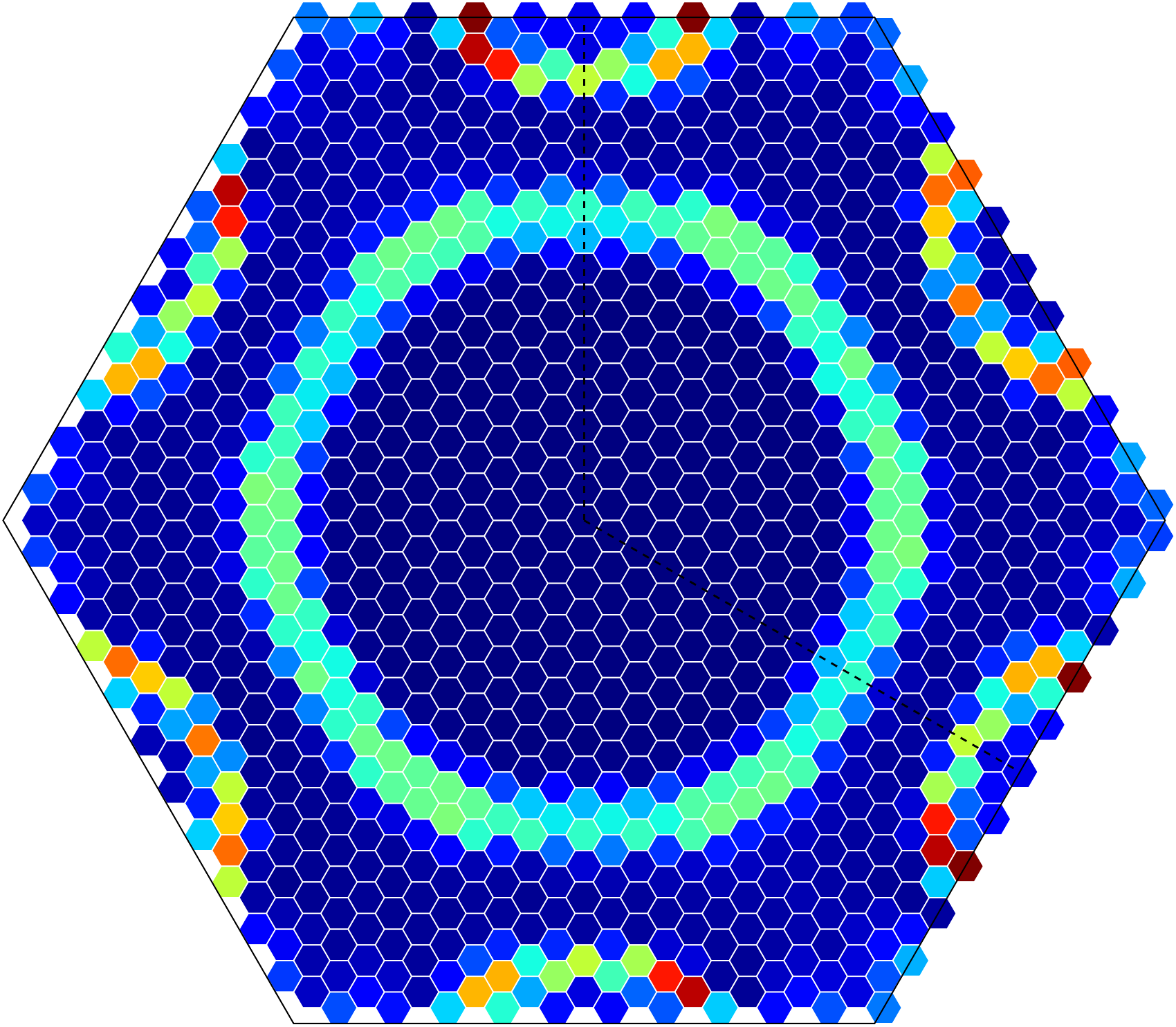}  \\
\includegraphics[width=\figwidth\textwidth]{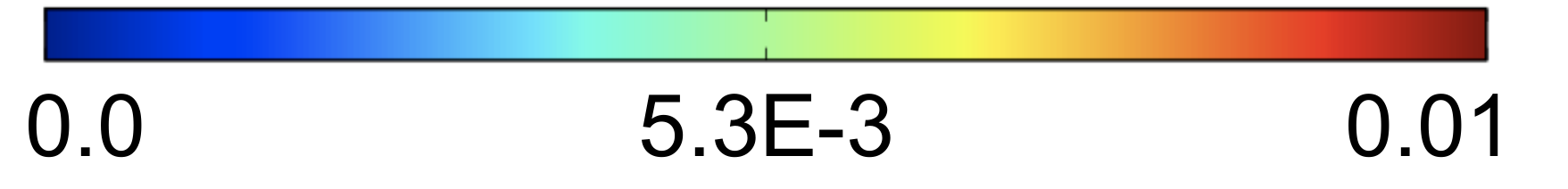}
\includegraphics[width=\figwidth\textwidth]{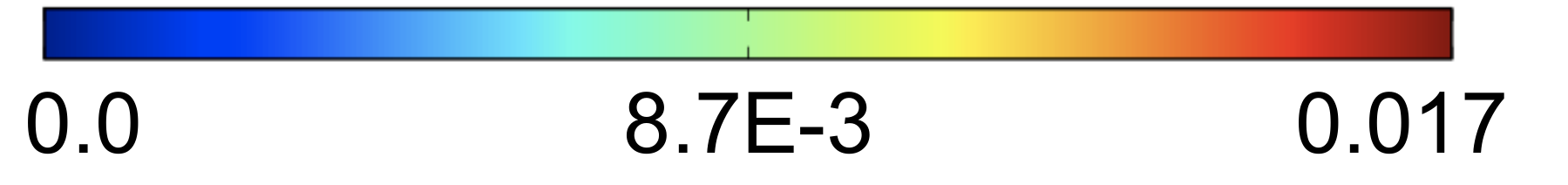} 
\includegraphics[width=\figwidth\textwidth]{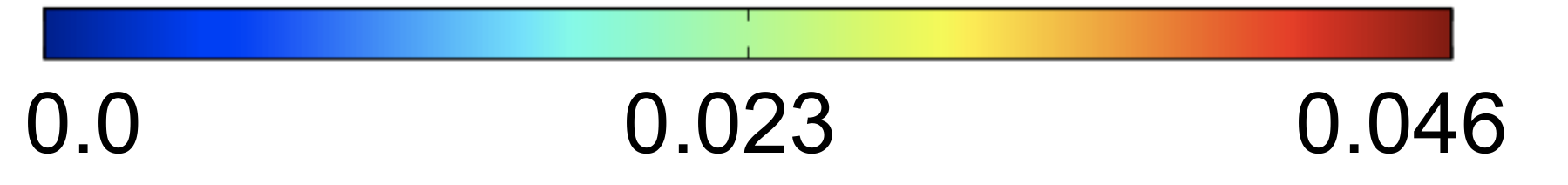} \\

\includegraphics[width=\figwidth\textwidth]{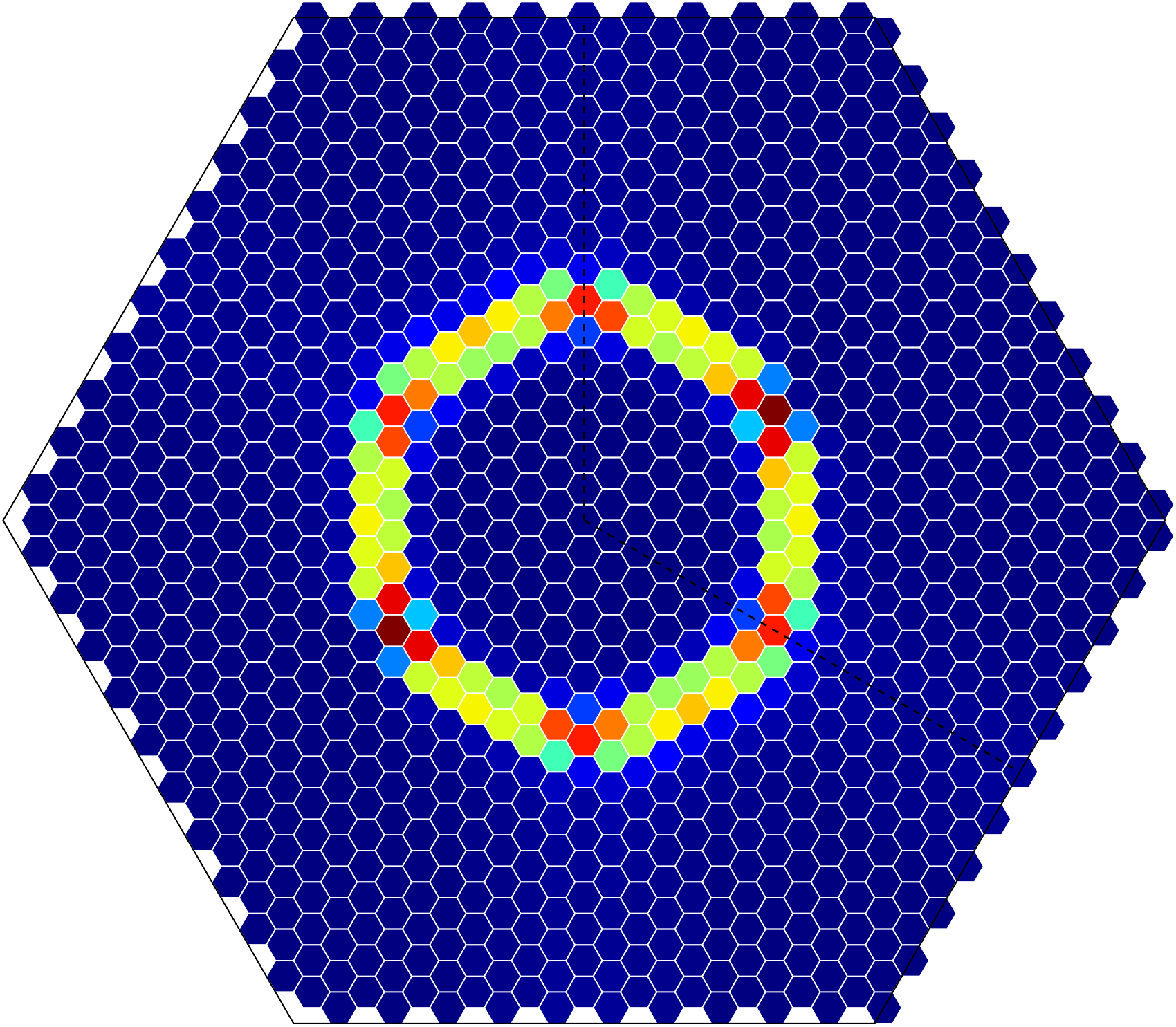} 
\includegraphics[width=\figwidth\textwidth]{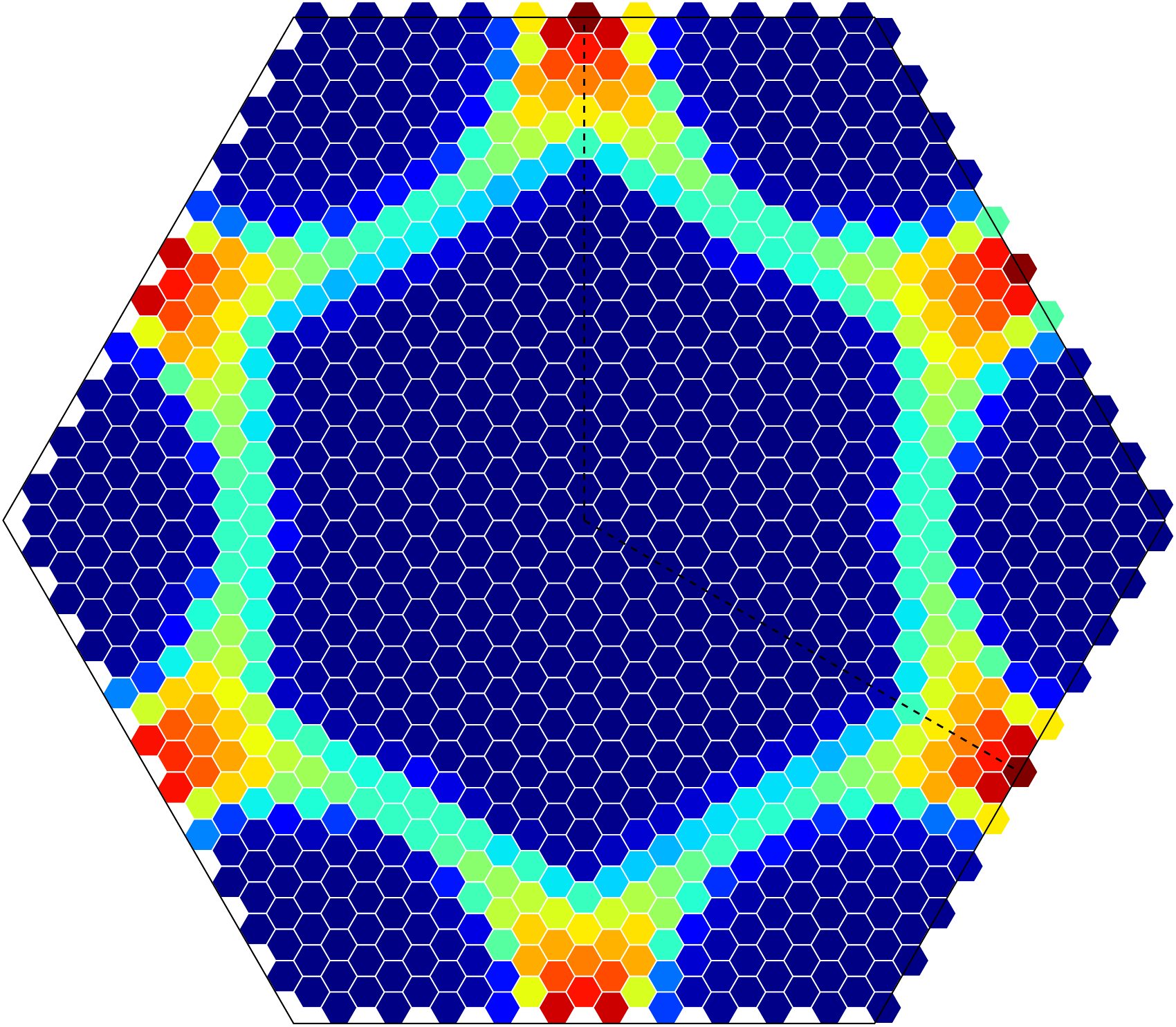} 
\includegraphics[width=\figwidth\textwidth]{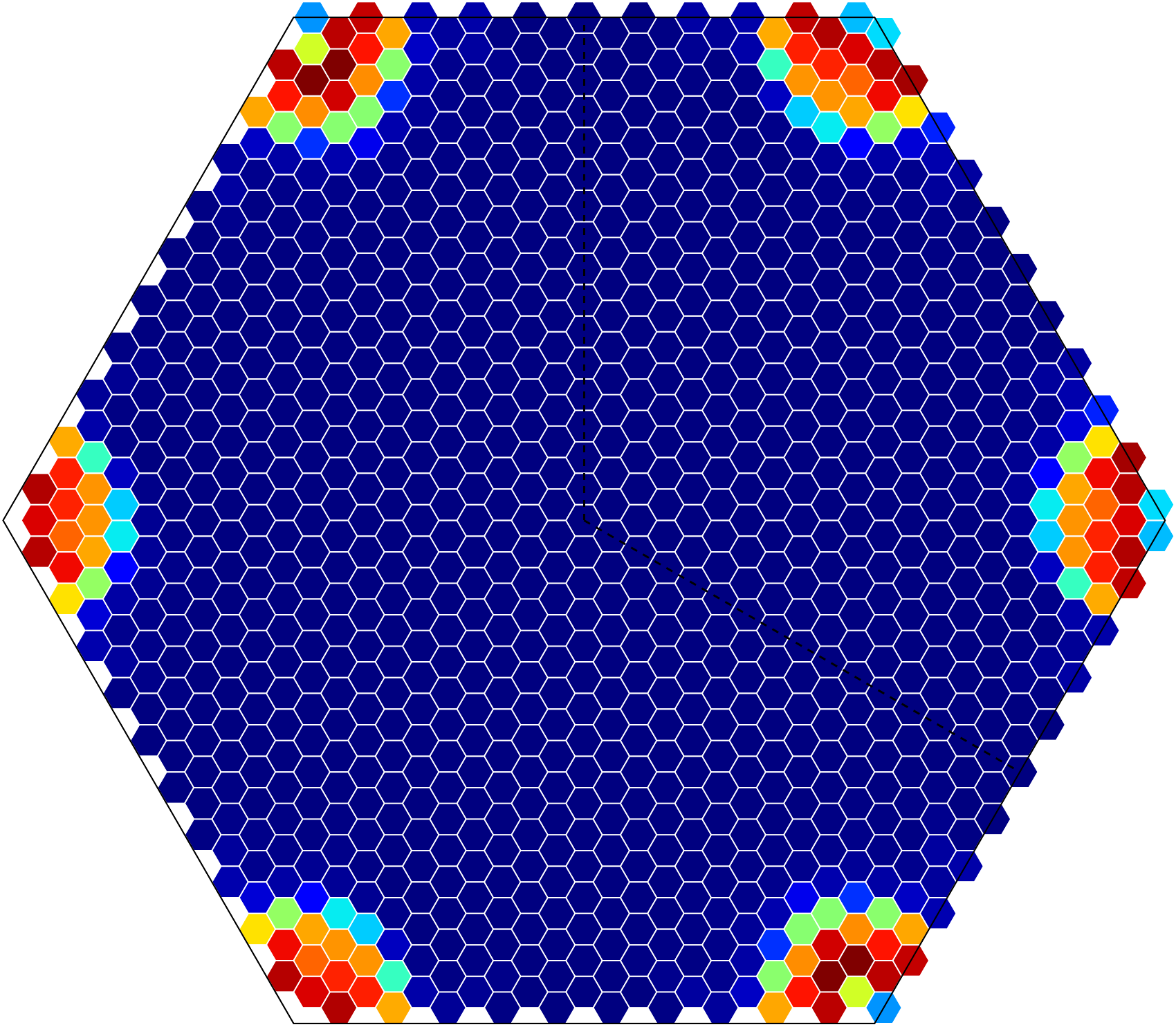}  \\
\includegraphics[width=\figwidth\textwidth]{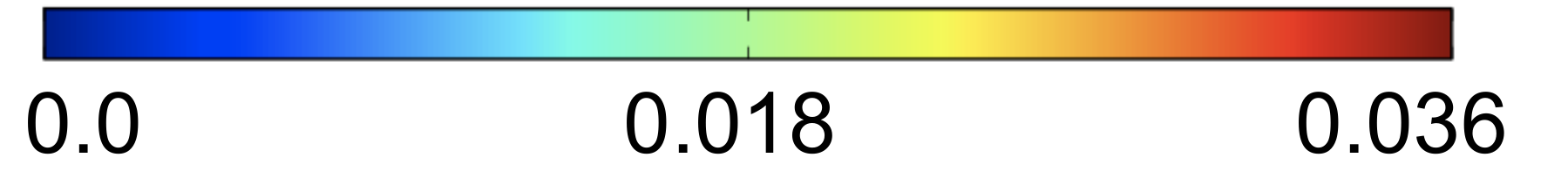}
\includegraphics[width=\figwidth\textwidth]{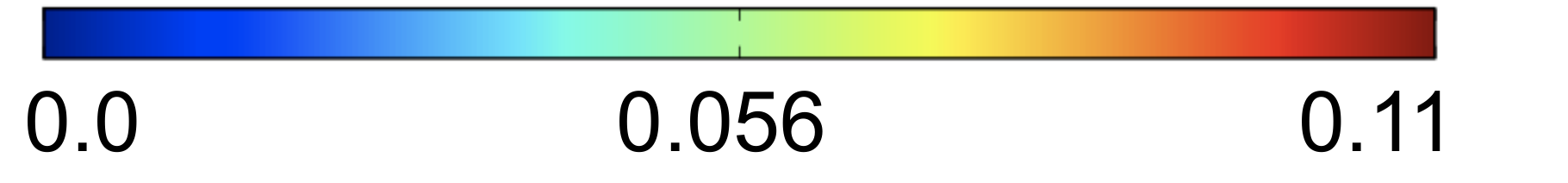}
\includegraphics[width=\figwidth\textwidth]{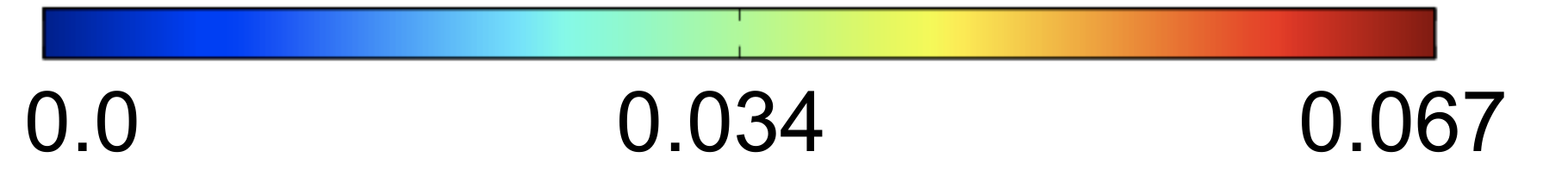} \\
\end{center}
\caption{
$S\big(\vec{q}, \omega\big)$ for a $(111)$ FCC patch ($P=32$, $C=32$) for various $\omega$. 
Upper Row: Longitudinal.
Lower Row: Transverse.
Left: $\omega{}/\omega_0=1.21$. Center: $\omega{}/\omega_0=1.65$. Right: $\omega{}/\omega_0=2.08$.
}
\label{fig: DSF 111 C32 P32}
\end{figure*}

\begin{figure*}[ph!]    
\begin{center}
\includegraphics[width=\figwidth\textwidth]{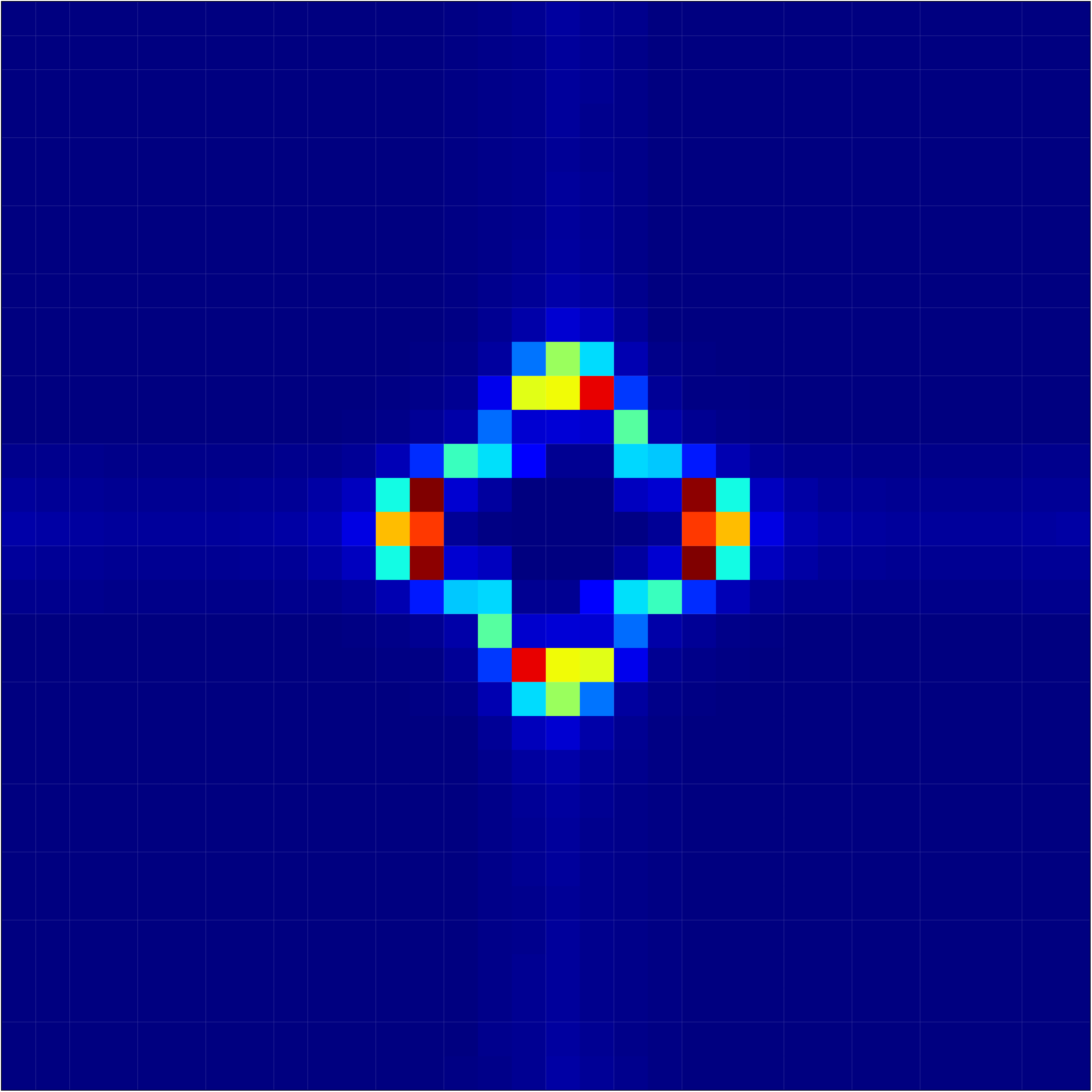} 
\includegraphics[width=\figwidth\textwidth]{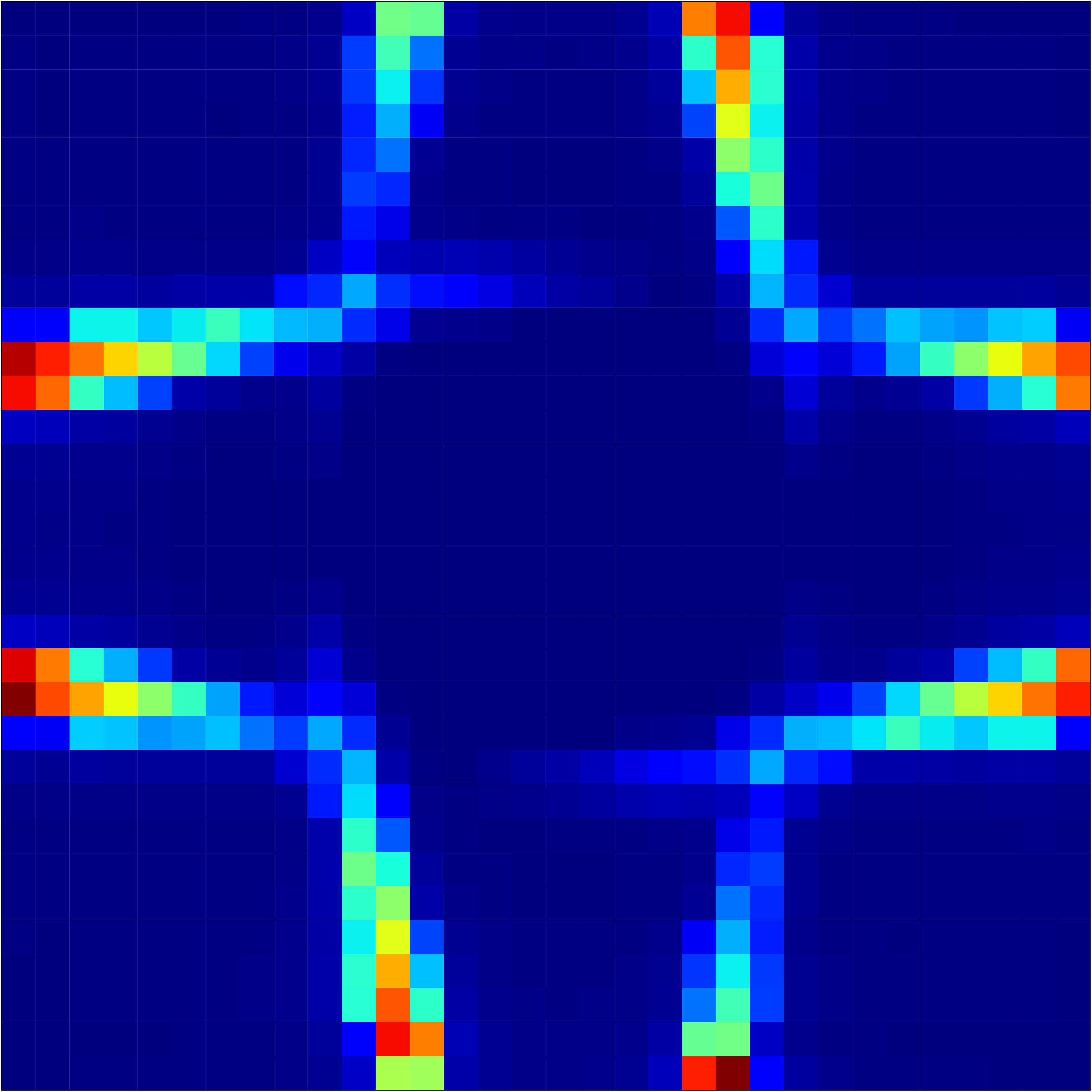} 
\includegraphics[width=\figwidth\textwidth]{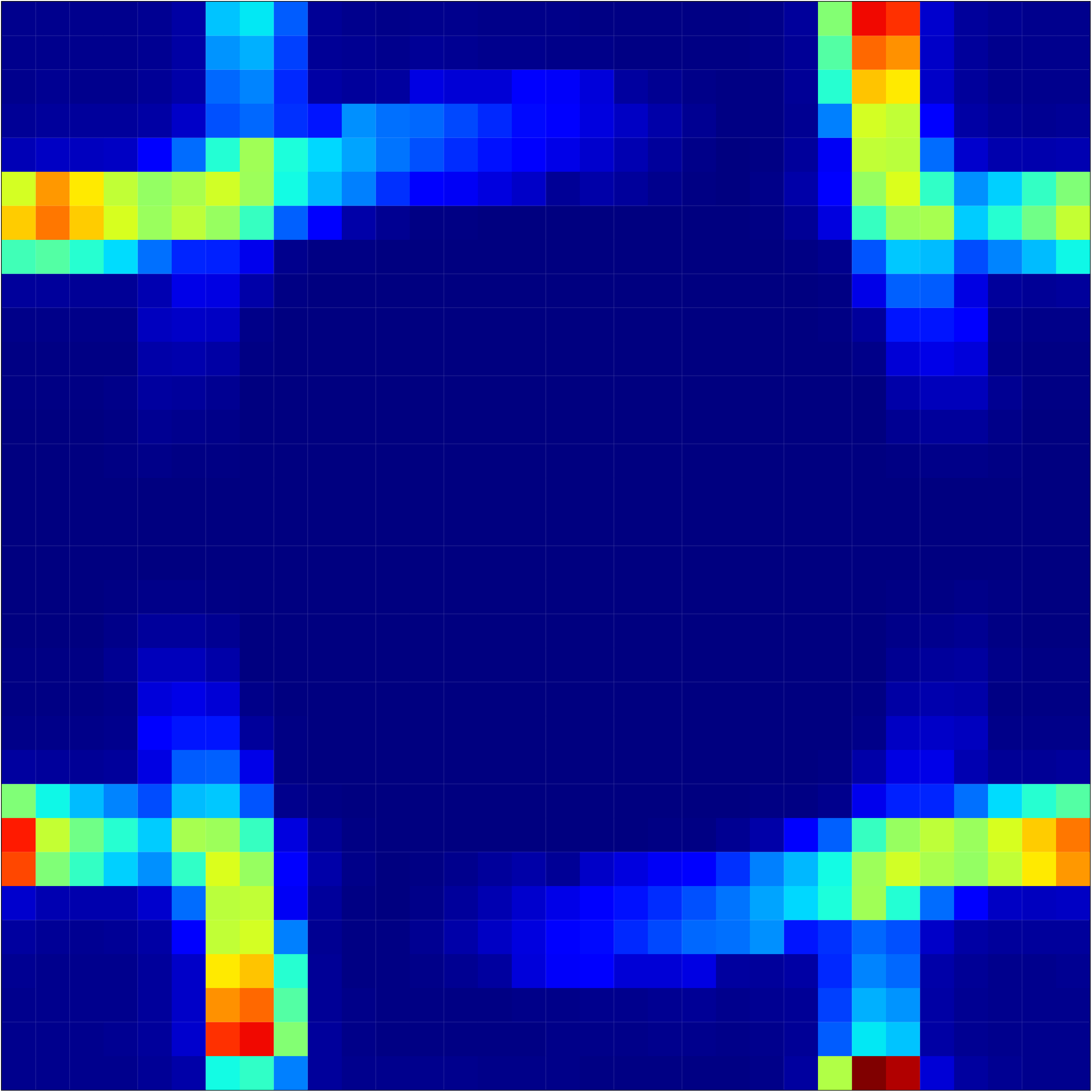}  \\
\includegraphics[width=\figwidth\textwidth]{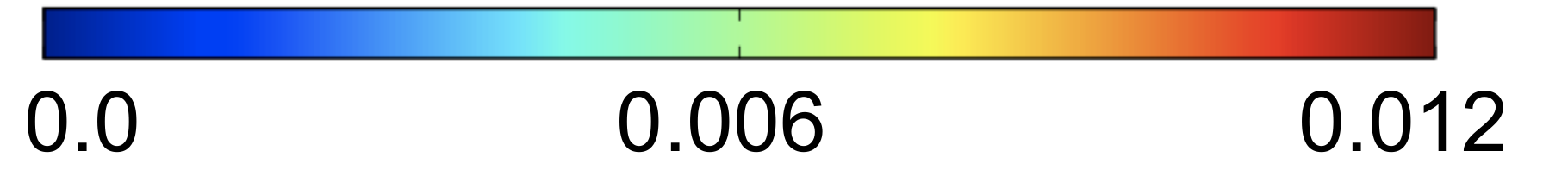}
\includegraphics[width=\figwidth\textwidth]{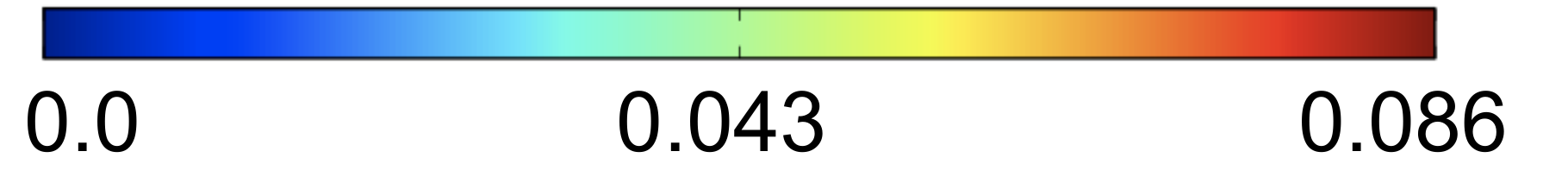} 
\includegraphics[width=\figwidth\textwidth]{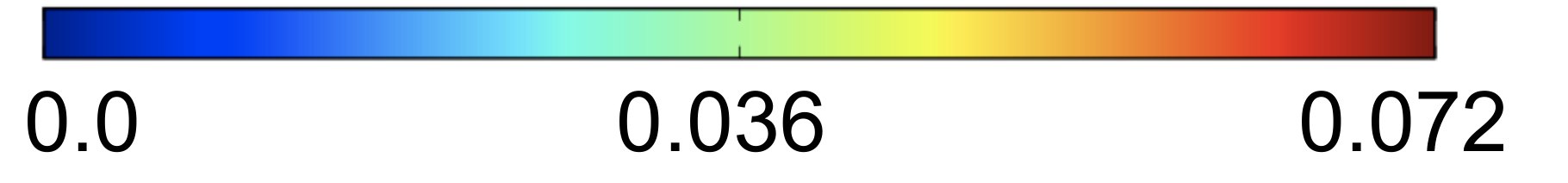} \\

\includegraphics[width=\figwidth\textwidth]{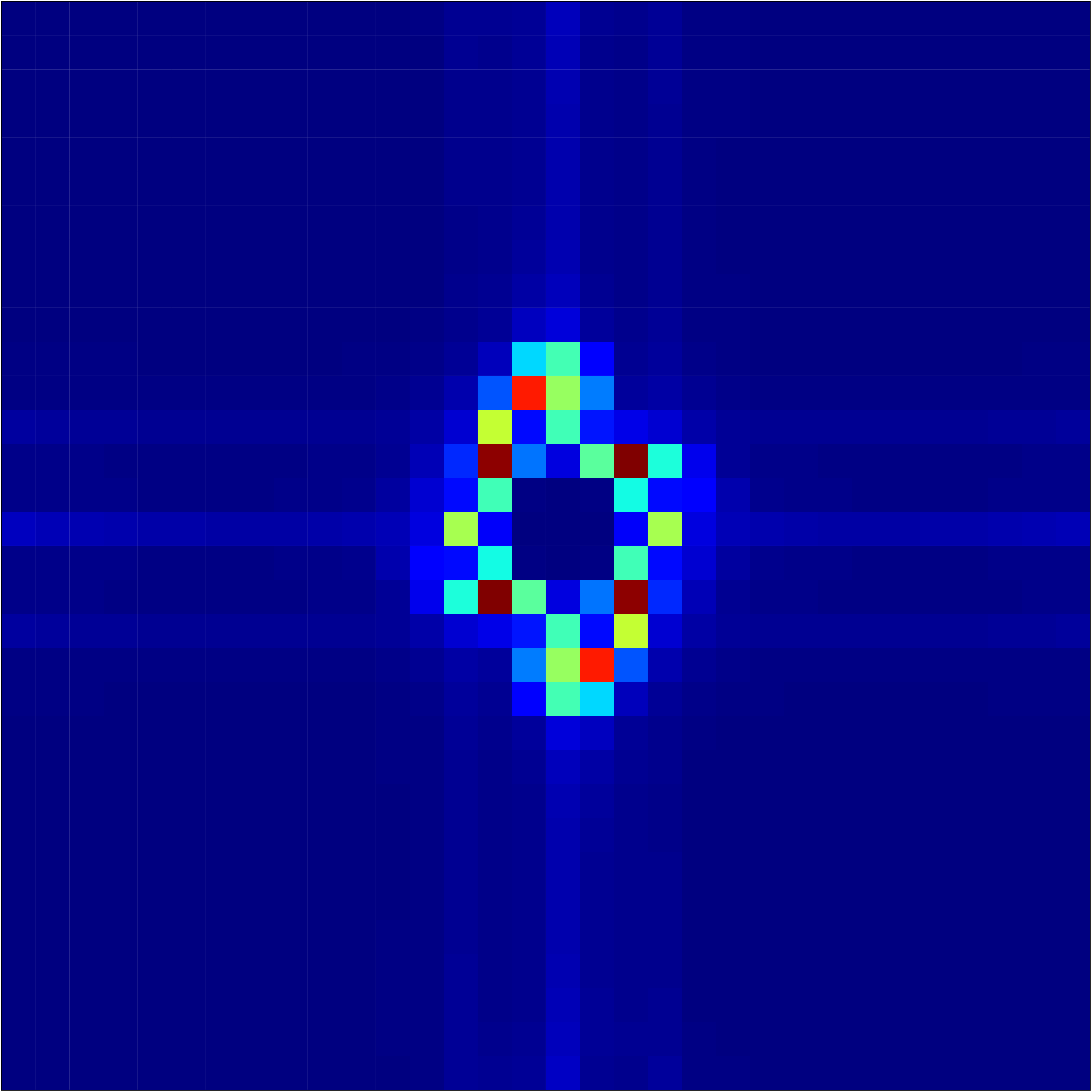} 
\includegraphics[width=\figwidth\textwidth]{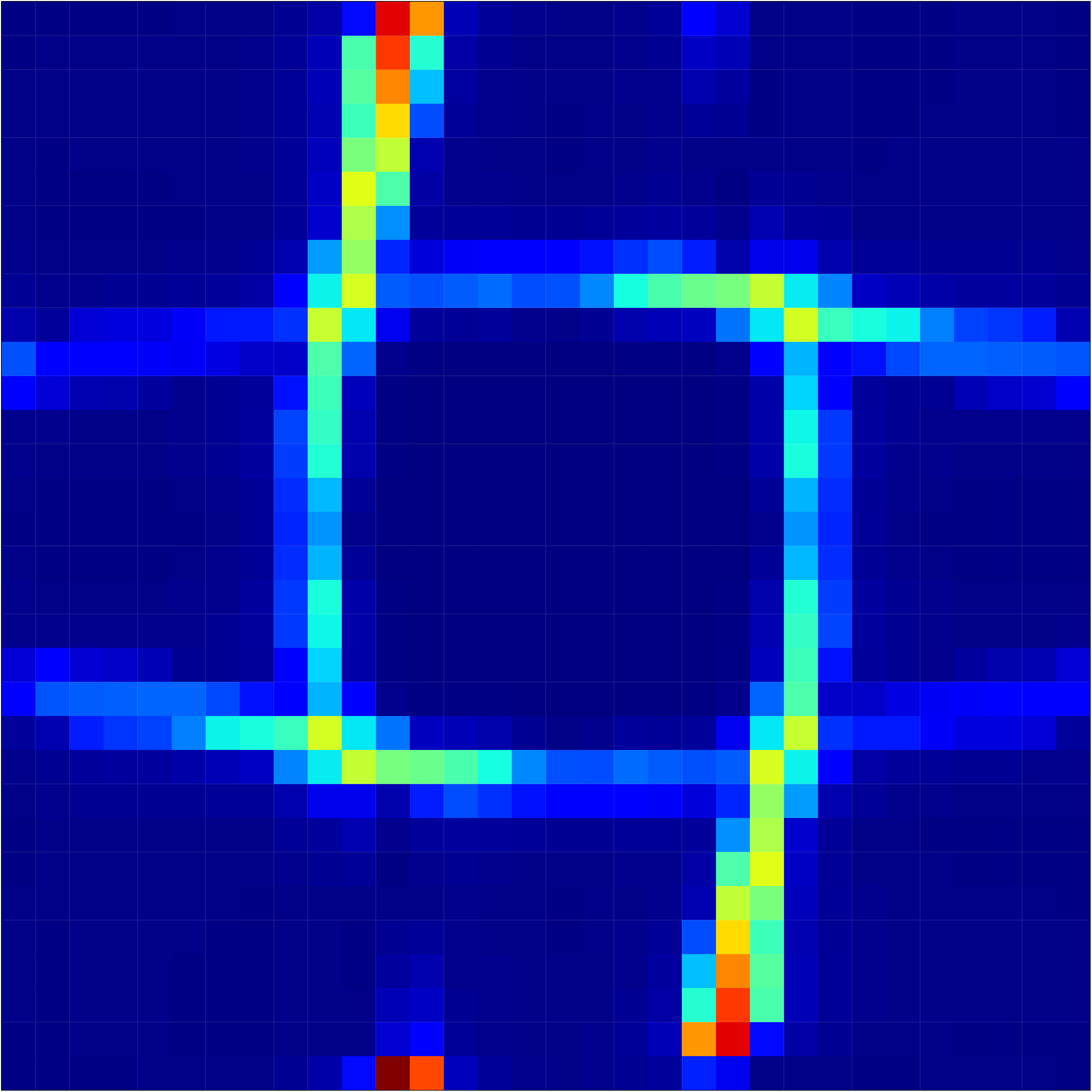} 
\includegraphics[width=\figwidth\textwidth]{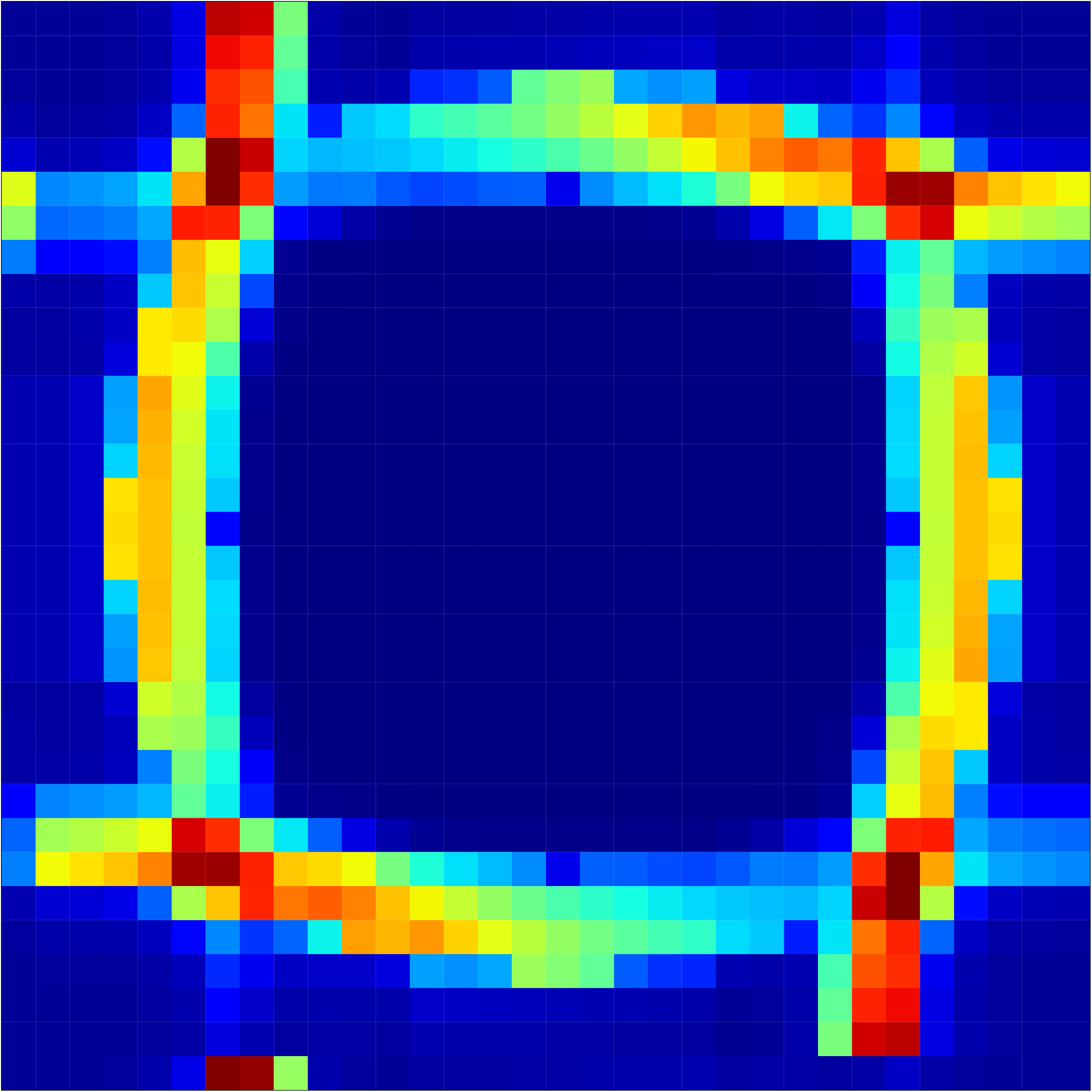}  \\
\includegraphics[width=\figwidth\textwidth]{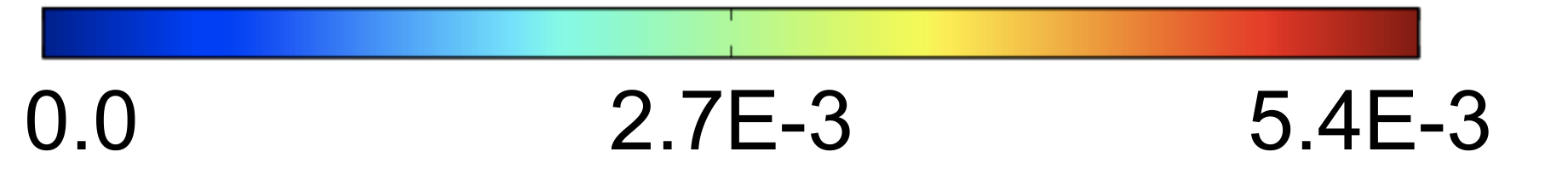}
\includegraphics[width=\figwidth\textwidth]{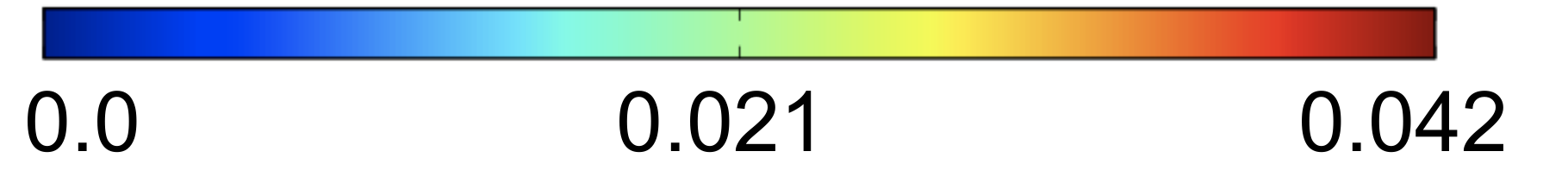}
\includegraphics[width=\figwidth\textwidth]{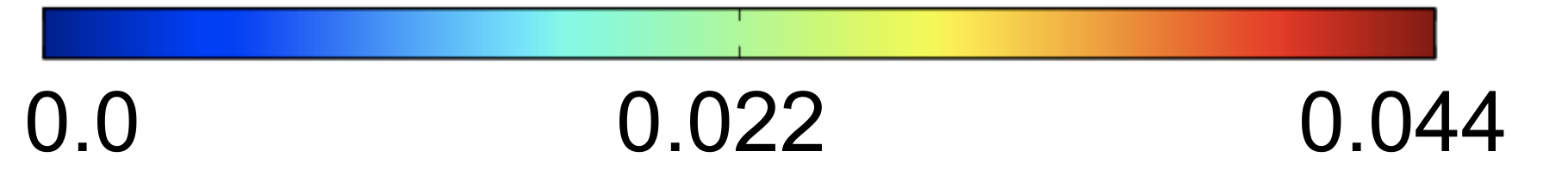} \\
\end{center}
\caption{
$S\big(\vec{q}, \omega\big)$ for a $(100)$ FCC patch ($P=32$, $C=32$) for various $\omega$. 
Upper Row: Longitudinal.
Lower Row: Transverse.
Left: $\omega{}/\omega_0=1.21$. Center: $\omega{}/\omega_0=1.65$. Right: $\omega{}/\omega_0=2.08$.
}
\label{fig: DSF 100 C32 P32}
\end{figure*}


We next focus our attention on the planewave decomposition of normal modes of the patch Green's function, $\mathcal{G}$, and their relationship to the dynamic structure factor (DSF)~\cite{AshcroftMermin}, the average square amplitude of plane waves in thermal equiliubrium.   
We start by computing the full set of normal modes $\hat{\phi}_{p}$, and eigenvalues $\lambda_{p}$ of $\mathcal{G}$  using the MATLAB eig() function.
$p$ indexes the modes.
We then perform a spatial Fourier transform of each normal mode and split into longitudinal and transverse components, respectively, as follows:  
\begin{equation}
\label{eqn: FT modes sq amps}
	\begin{aligned}
	E_{L}^{p}(\vec{q})=\bigg|\sum_{\vec{r}} \big[\hat{q}\cdot\hat{\phi}_{p}(\vec{r})\big]e^{i\dotP{q}{r}} \bigg|^{2} \\
	E_{T}^{p}(\vec{q})=\bigg|\sum_{\vec{r}} \big[\hat{q}\times\hat{\phi}_{p}(\vec{r})\big]e^{i\dotP{q}{r}} \bigg|^{2}
	\end{aligned}
\end{equation}
Here $\vec{r}$ labels a lattice site; $E_{L}^{p}(\vec{q})$ and $E_{T}^{p}(\vec{q})$ are square amplitudes of the longitudinal and transverse Fourier transform components, respectively, for a reciprocal lattice vector (of the patch) $\vec{q}$.

Next we compute the longitudinal (transverse) DSF, $S_{L(T)}$, which is written in terms of normal modes as~\cite{Schober:2004JPhysCondMat, Shintani:2008natMat}:
\begin{equation}
\label{eqn: DSF definition}
S_{L(T)}\big(\vec{q}, \omega\big) = \frac{k_{b}Tq^{2}}{m\omega^{2}}\sum_{p} E_{L(T)}^{p}(\vec{q})\Theta\left(0.1 - \bigg| \frac{\omega - \omega_{p}}{\omega} \bigg|\right)
\end{equation} 
Here $m$ is particle mass and the temperature is such that $k_{b}T/Ka^2=1$.
The sum runs over modes indexed by $p$ with corresponding energy eigenfrequency $\omega_{p}=1/\sqrt{m\lambda_{p}}$.
$\Theta\left(x\right)$ is the Heaviside step function, which is equal to $1$ when $x>0$ and $0$ otherwise.
$S\big(\vec{q}, \omega\big)$, encodes the participation of various planewaves in a normal mode of eigenfrequency $\omega$~\cite{Kaya:2011pre}.

Figures~\ref{fig: DSF 111 C32 P32} and~\ref{fig: DSF 100 C32 P32} show $S\big(\vec{q}, \omega\big)$ plotted in the first BZ for $(111)$ and $(100)$ FCC patches for a set of 3 values of $\omega$ that are typical low, mid-range and high eigenfrequencies of $\mathcal{G}$.
We chose these frequencies so that the `high' frequency is roughly near the second van Hove peak in the DOS,  while the `mid' frequency is near the first van Hove peak and the `low' frequency is lower in the DOS. 
In both plots the upper row shows the longitudinal component of the DSF while bottom row shows the transverse component.

The $(111)$ patch DSF plots of figure~\ref{fig: DSF 111 C32 P32} show that low energy normal modes (left column) have dominant participation from planewaves of a single wavenumber.
The longitudinal planewaves participate at roughly half the wavenumber as transverse planewaves  in modes of a given frequency $\omega$ and the DSF is isotropic as we expect from the low $k$ dispersion of a $(111)$ patch.
For modes of intermediate frequency (middle column) which correspond to the first van Hove peak in the DOS (see bottom plot of  figure~\ref{fig: DOS computed}), the dominant transverse planewaves are near the Brillouin Zone boundary. 
The DSF for higher frequency modes (right column) has peaks at multiple wavenumbers, however participating planewaves are still localized in the BZ.
The dominant longitudinal planewaves are near the BZ boundary, as one would expect since $\omega$ is near the second van Hove peak. 
In addition, while the DSF has hexagonal symmetery, the longitudinal component is fairly isotropic except for the highest frequency normal modes.
This is consistent with the pesudo-dispersion of $\mathcal{G}$ of $(111)$ patches seen in figure~\ref{fig: complete C32 [111] analytical dispersion}.

In contrast, the DSF for the $(100)$ patch (figure~\ref{fig: DSF 100 C32 P32}) shows participation from planewaves of different wavenumber in a given narrow frequency range. 
Roughly speaking, at a given $\omega$, the DSF is peaked at the equi-$\omega$ contours of the dispersion plot.
Much like the dispersion of $(100)$ patches, the DSF is highly anisotropic.
At frequencies below the first van Hove singularity in the DOS (top plot of figure~\ref{fig: DOS computed}) the participating planewaves in the DSF are of relatively low wavenumber (left plot), whereas even for modes with frequency just beyond the first van Hove singularity we see dominant participating transverse \emph{and} longitudinal planewaves near the BZ boundary (middle and right column).
 
We also compute an isotropic average of $S(\vec{q}, \omega)$ in wavevector space.
\begin{equation}
\label{eqn: isotropic DSF definition}
S_{L(T)}\big(q, \omega\big) =\frac{1}{N_{q}} \sum_{\vec{q}} S_{L(T)}(\vec{q}, \omega)\Theta\left(0.1 - \bigg| \frac{q - \big|\vec{q}\big|}{q} \bigg|\right)
\end{equation} 
\begin{equation}
N_{q}=\sum_{\vec{q}} \Theta\left(0.1 - \bigg| \frac{q - \big|\vec{q}\big|}{q} \bigg|\right)
\end{equation}

Figures~\ref{fig: isotropic DSF 111 C32 P32} and~\ref{fig: isotropic DSF 100 C32 P32} show $S\big(q, \omega\big)$ computed for the Green's function of a $(111)$ and a $(100)$ patch, respectively. 
In each plot, the vertical axis spans the range of frequencies of $\mathcal{G}$ while the horizontal axis spans the range of wavenumbers in the first BZ of the patch. 

\newcommand{\figsize}{0.235}

\begin{figure}[t]   
\begin{center}
\includegraphics[width=\figsize\textwidth]{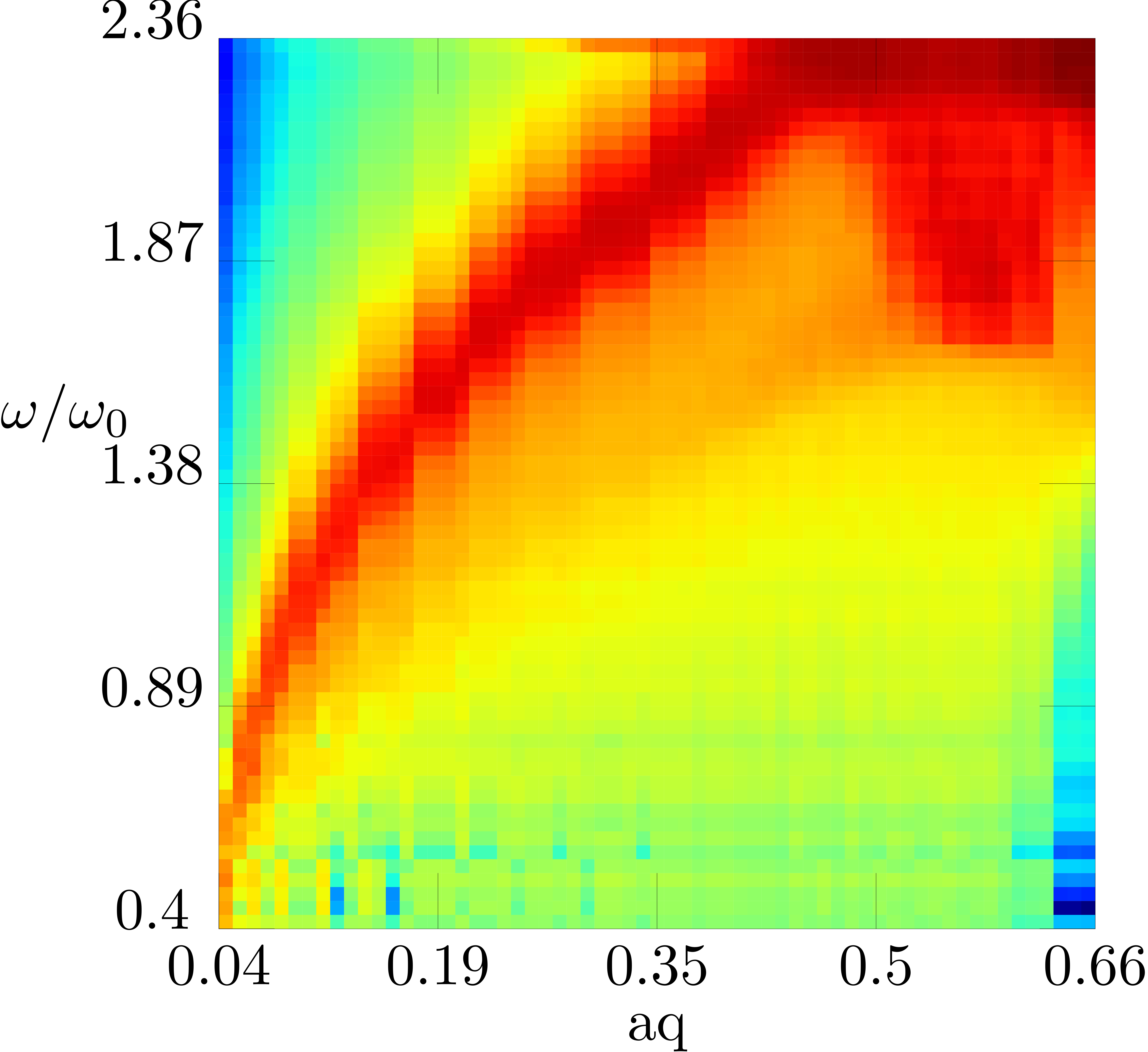} 
\includegraphics[width=\figsize\textwidth]{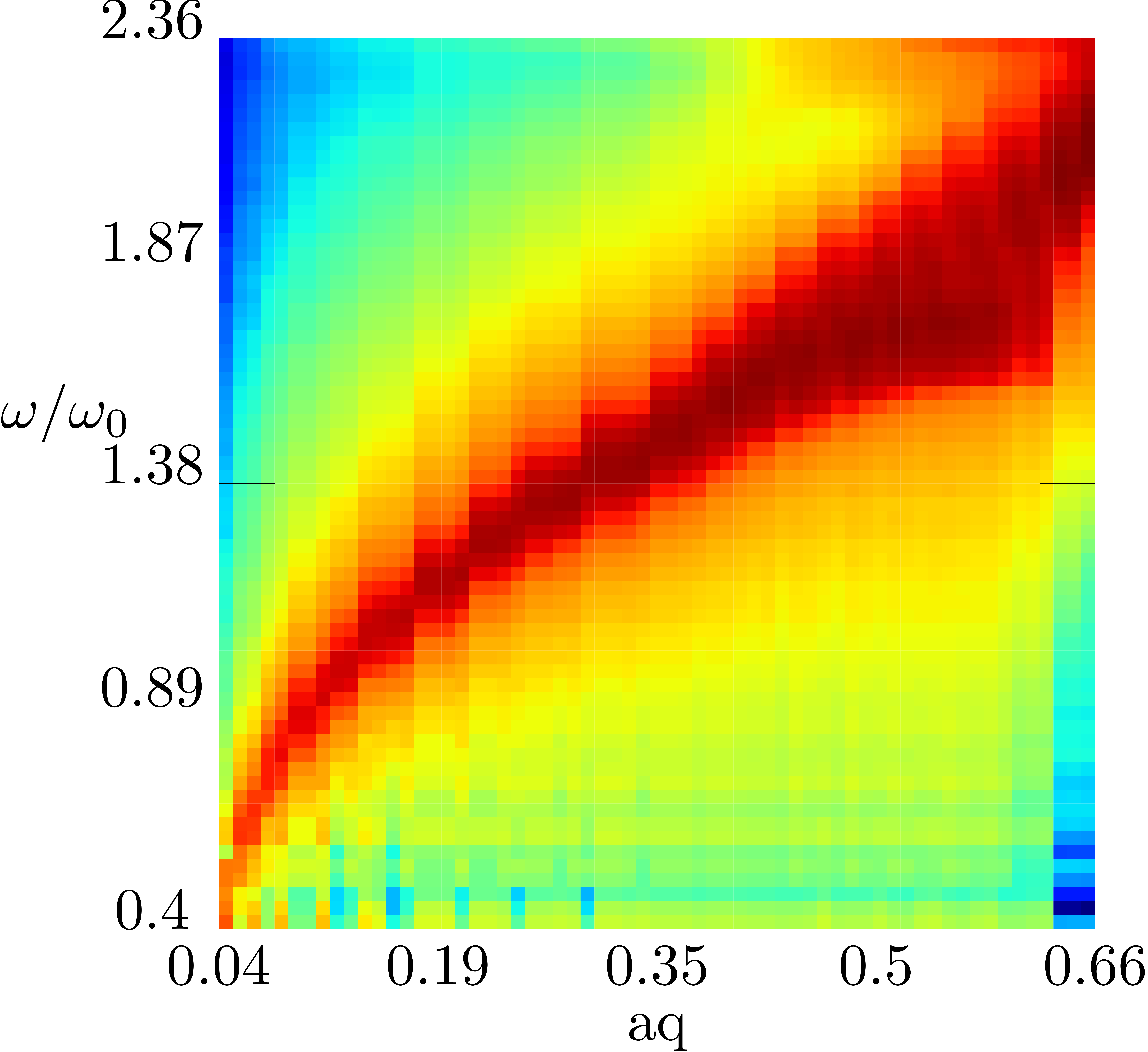} 
\includegraphics[width=\figsize\textwidth]{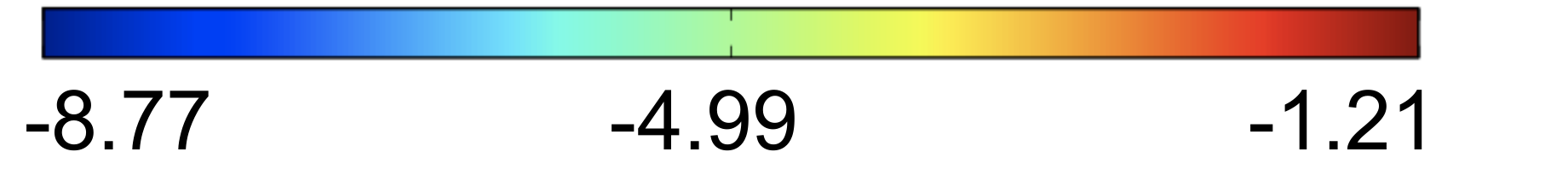} 
\includegraphics[width=\figsize\textwidth]{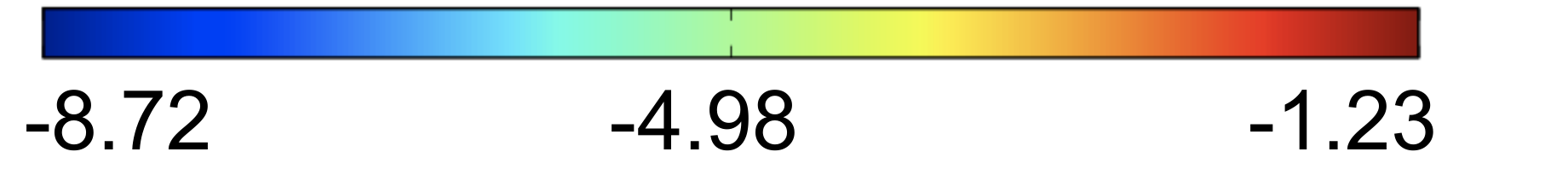}
\end{center}
\caption{
Isotropically averaged DSF $S\big(q, \omega\big)$, for a $(111)$ FCC patch ($P=32$, $C=32$).
The colorbar is {\it log scale}.
Left: Longitudinal.
Right: Transverse.
}
\label{fig: isotropic DSF 111 C32 P32}
\end{figure}

\begin{figure}[t]    
\begin{center}
\includegraphics[width=\figsize\textwidth]{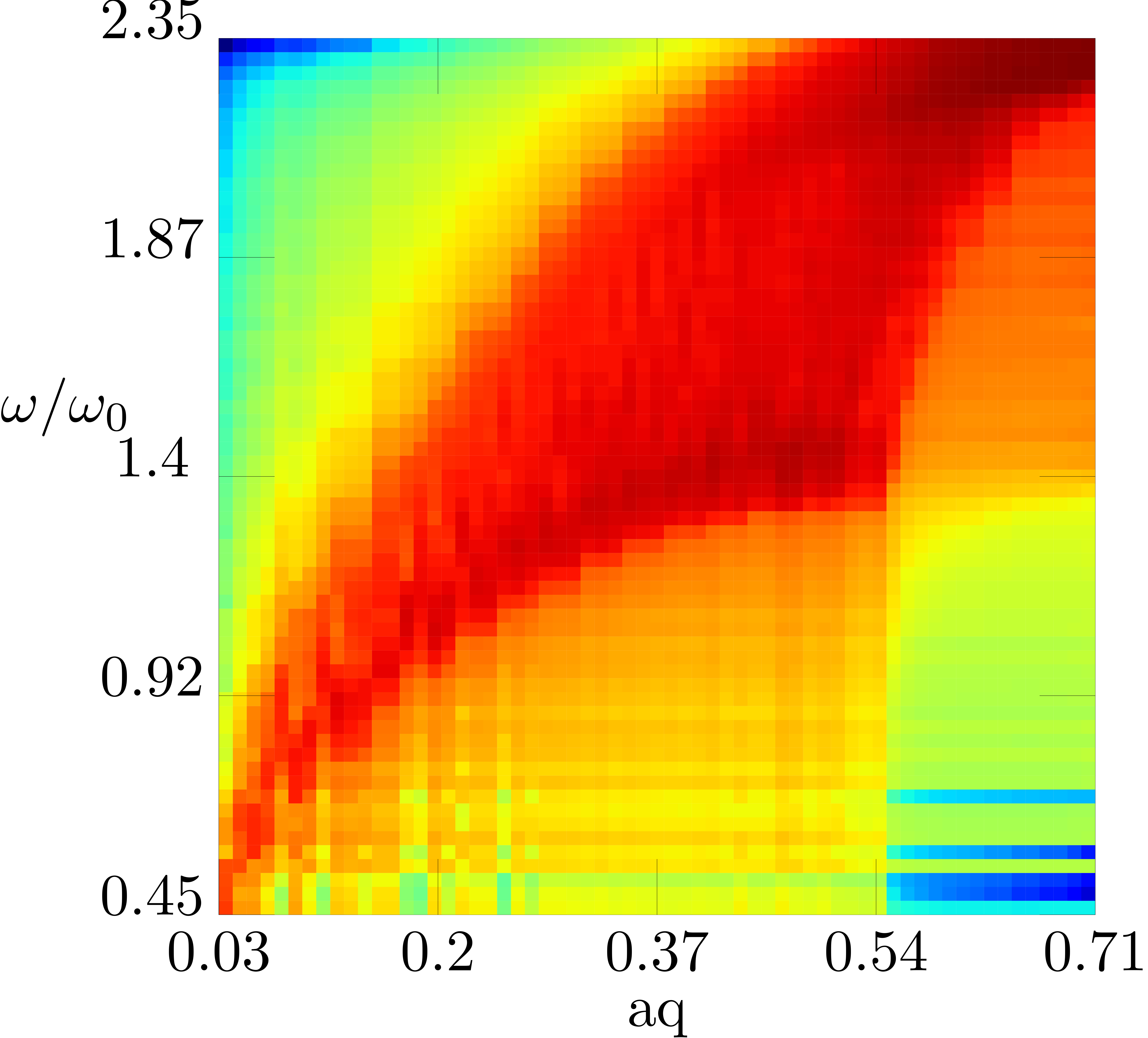} 
\includegraphics[width=\figsize\textwidth]{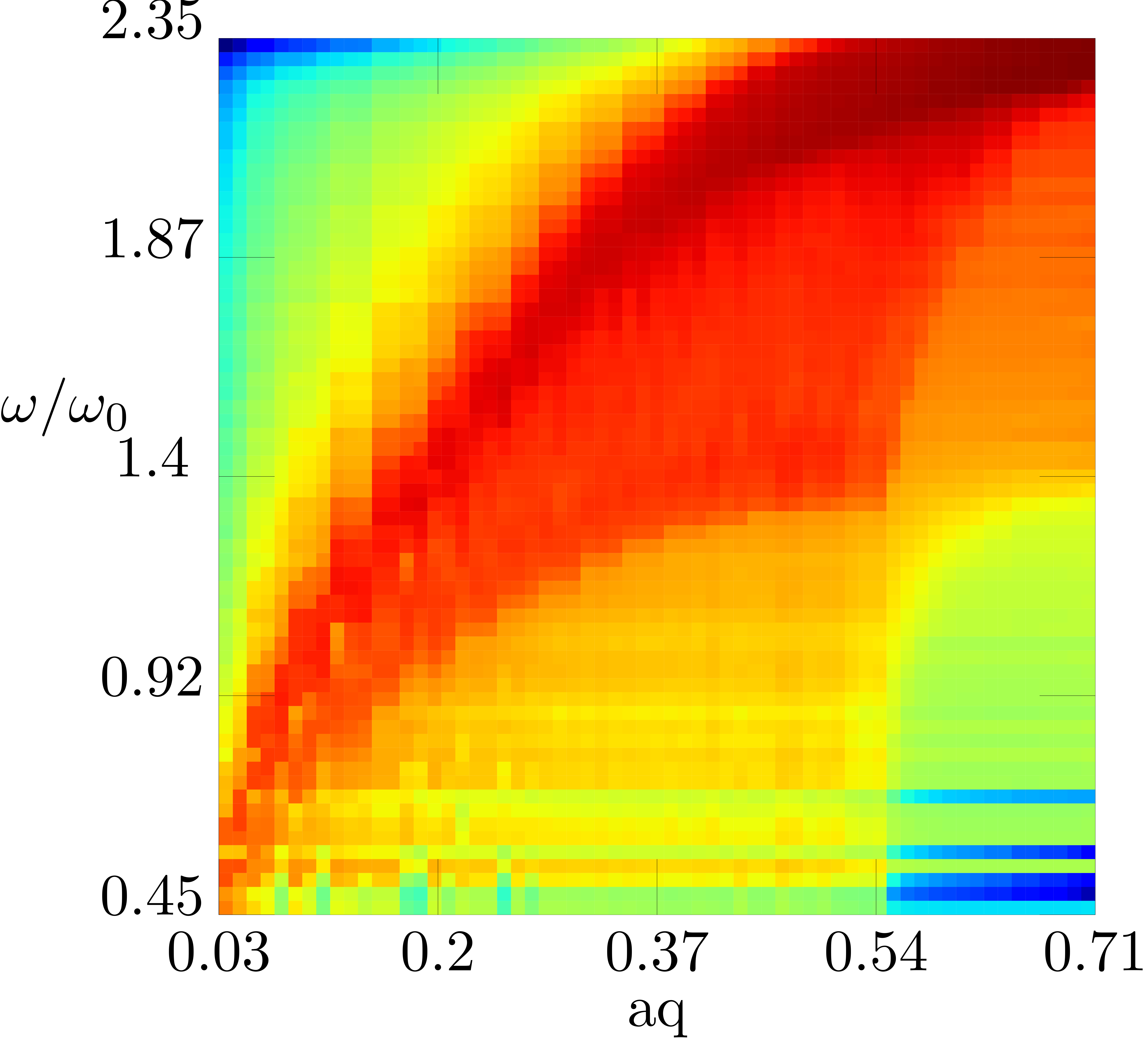} 
\includegraphics[width=\figsize\textwidth]{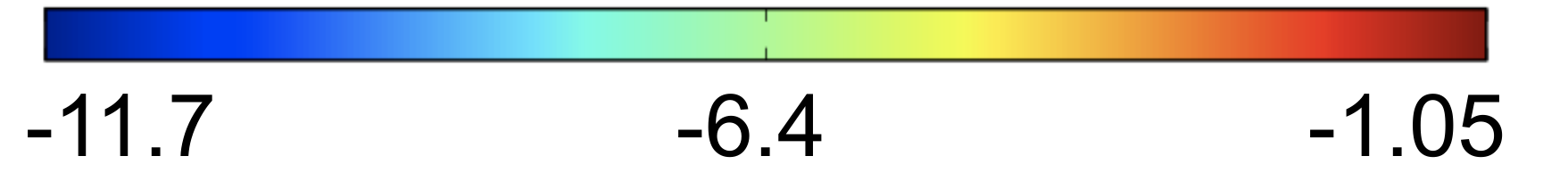} 
\includegraphics[width=\figsize\textwidth]{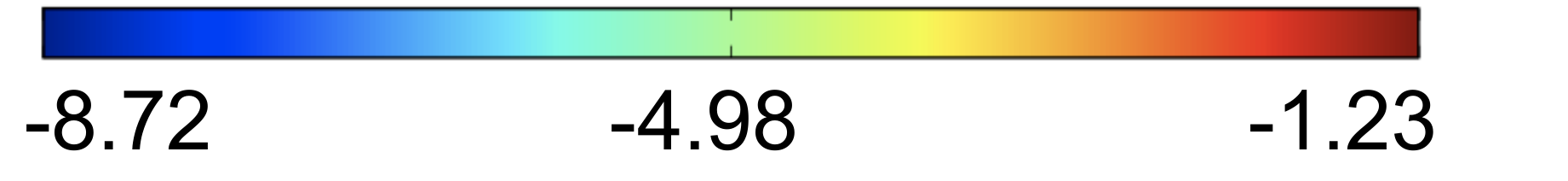}
\end{center}
\caption{
Isotropically averaged DSF $S\big(q, \omega\big)$ for a $(100)$ FCC patch ($P=32$, $C=32$).
The colorbar is {\it log scale}.
Top: Longitudinal.
Bottom: Transverse.
}
\label{fig: isotropic DSF 100 C32 P32}
\end{figure}


In figure~\ref{fig: isotropic DSF 111 C32 P32} we see that for fixed, lower values of $\omega$, $S\big(q, \omega\big)$ has a peak at a single value of $q$.
Low energy modes of $\mathcal{G}$ for $(111)$ patches have dominant participation by a planewaves of a single $q$.
In contrast, high energy modes have multiple contributing planewaves.
For example, modes around $\omega{}/\omega_{0}=2.08$, $S_{L}\big(\vec{q}, \omega\big)$ (upper plot, right column of figure~\ref{fig: DSF 111 C32 P32}) have peaks for at least 2 values of $q$.
As we expect from the pseudo-dispersion in figure~\ref{fig: complete C32 [111] analytical dispersion}, a longitudinally polarized planewave of wavenumber $q$ participates in a normal mode of higher $\omega$ than a transverse planewave.
In contrast to the $(111)$ case, the isotropically averaged $(100)$ DSF contains a wide range of $q$ values at a given $\omega$.
This can be understood from the highly anisotropic dispersion.




\section{Monte-Carlo simulation}
\label{section: Monte-Carlo simulation}

Experimental studies have access to only finite statistical samples and will be subject to artifacts arising from incomplete statistical information.
To study these artifacts, we perform Monte-Carlo (MC) simulations using \emph{precisely} the same Hamiltonian as our analytical calculations.
In particular we focus on the convergence of DOS.
 
Using the Metropolis algorithm, we perform MC simulations to sample configurations from a Gibbs-Boltzmann distribution.
Our system is a cubic FCC crystal with periodic boundary conditions, having $C=32$ cubic unit cells along each edge ($4\times{}32^{3}$ particles). 
The crystal Hamiltonian is constrained to be harmonic by specifying a Hessian $H_{i\alpha{}j\beta}$ ($i,j$  label atoms while $\alpha, \beta$ label Cartesian axes) that includes only nearest-neighbor interactions.
The energy is quadratic in particle displacements $\{x_{i\alpha}\}$, and is given by: $E(\{x_{i\alpha}\}) = \frac{1}{2}\sum_{i\alpha{}j\beta}x_{i\alpha}H_{i\alpha{}j\beta}x_{j\beta}$. 
We work in the same units as before: energy is measured in units of $Ka^2$.
The temperature is set so that $k_{b}T/Ka^2=1$.
We use MC steps whose size is sampled randomly from a uniform distribution of width $0.95a$, where $a$ is the nearest spacing.
This gives us an acceptance ratio of approximately $81\%$.   

\begin{figure}[t]    
\begin{center}
\includegraphics[width=.49\textwidth]{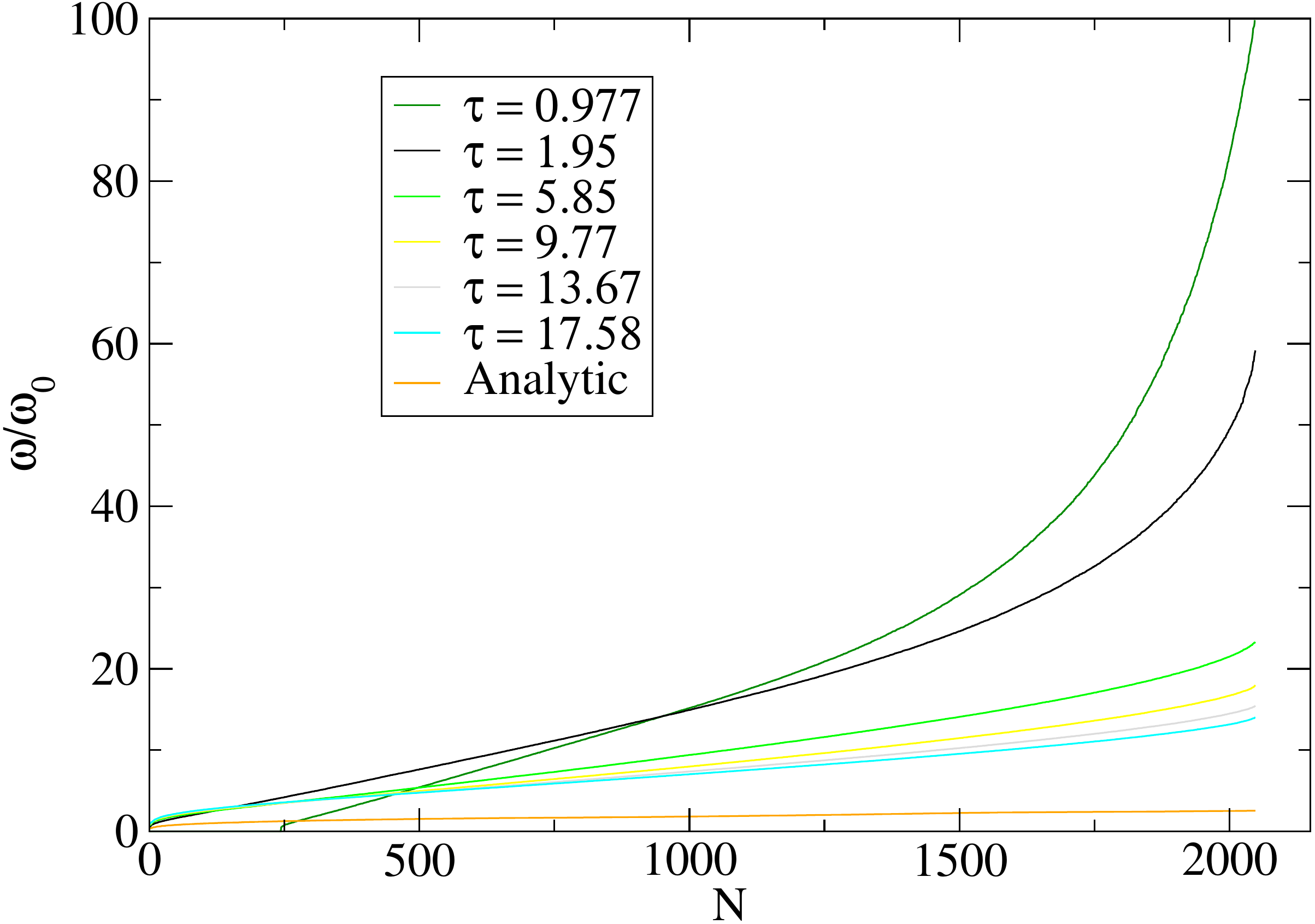}
\includegraphics[width=.49\textwidth]{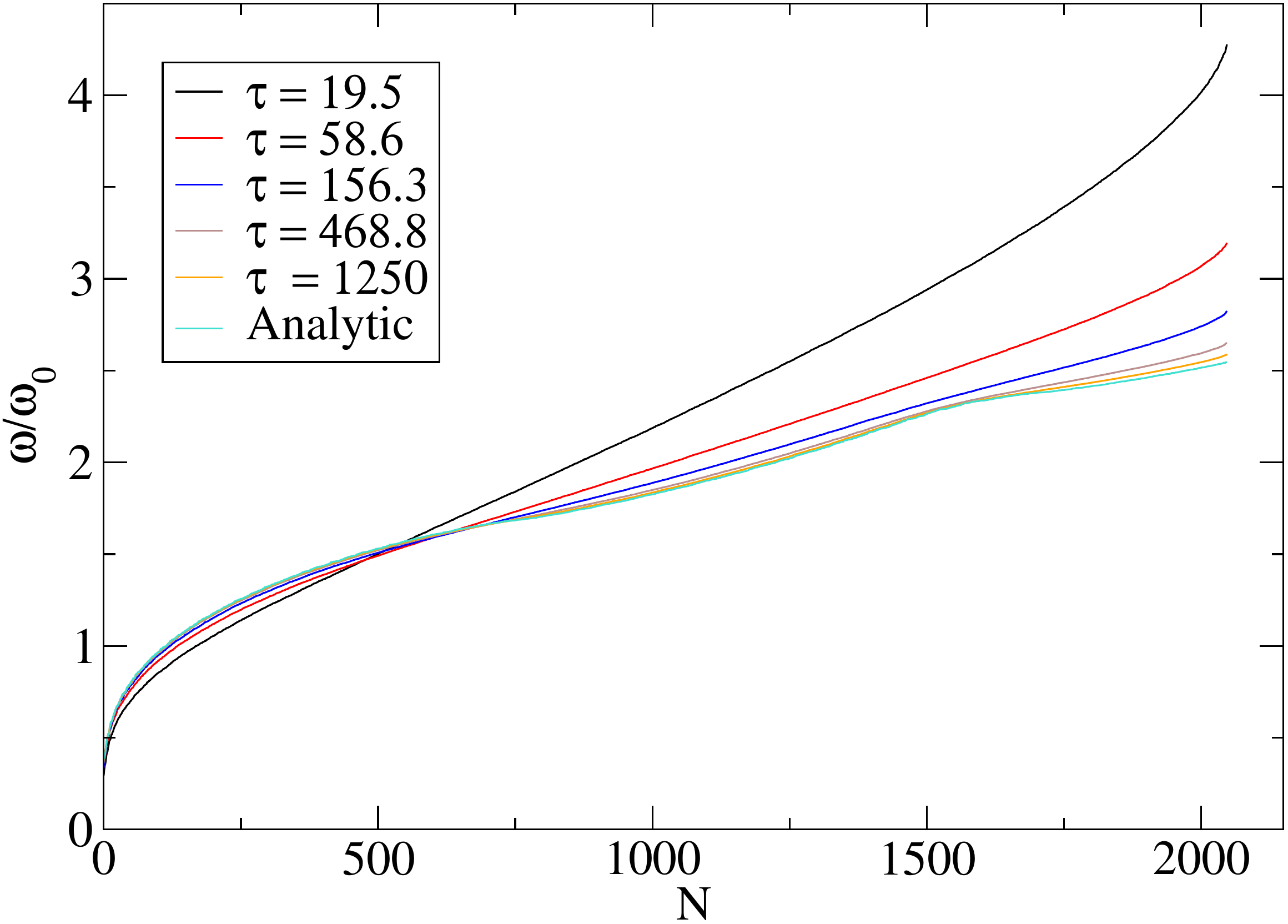}  
\end{center}
\caption{
Evolution of the frequency spectrum of the Green's function of a $(111)$ FCC patch ($P=32$, $C=32$) inferred from Monte-Carlo simulations.
 $\tau$ is the number of MC sweeps per patch degree of freedom. 
Top: Small sample.
Bottom: Longer sample.
}
\label{fig: patch MC eigenvalue evolution}
\end{figure}

\begin{figure}[t!]    
\begin{center}
\includegraphics[width=.49\textwidth]{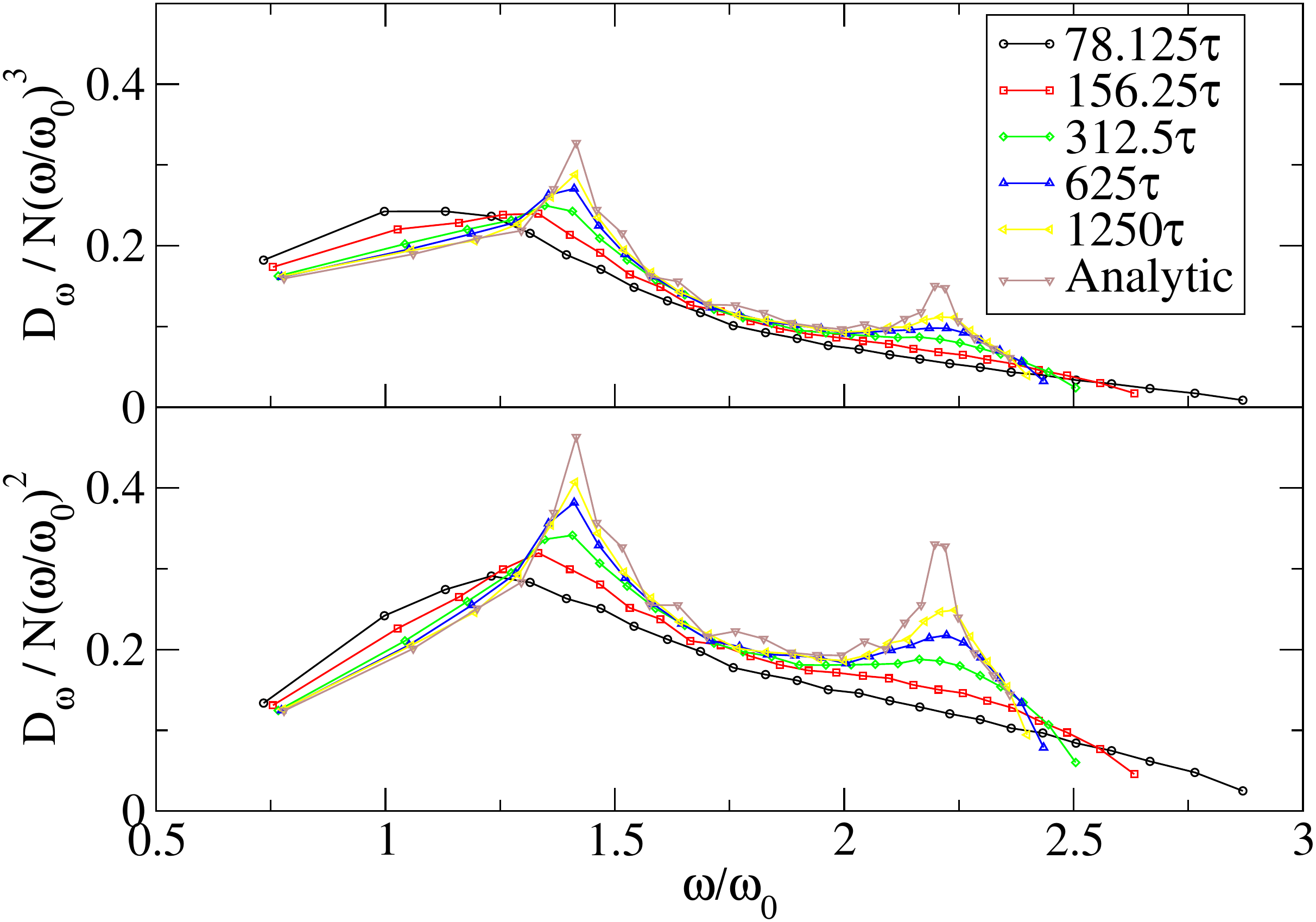}
\includegraphics[width=.49\textwidth]{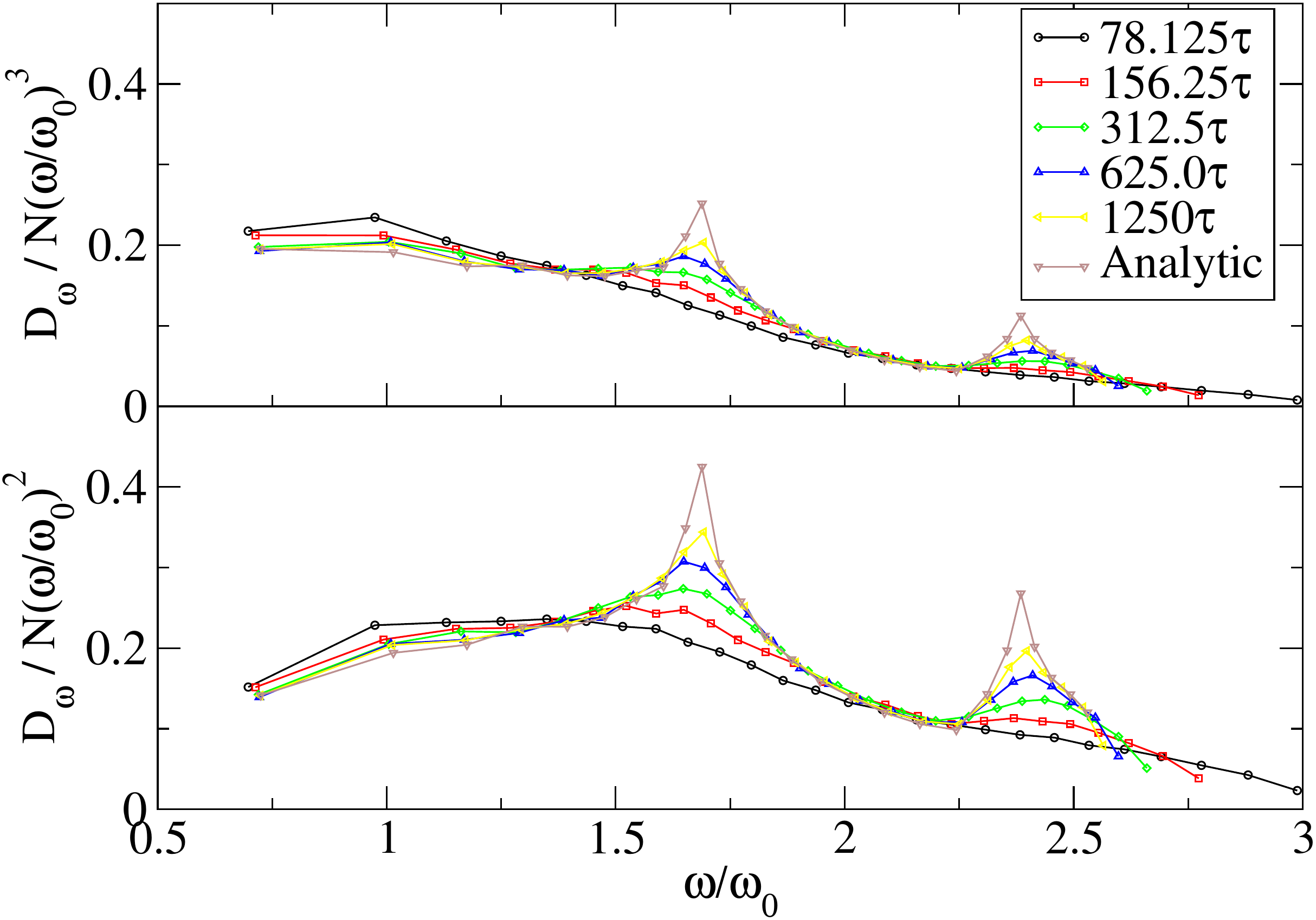}
\end{center}
\caption{
Scaled DOS, $D_{\omega}$, of a FCC crystal patch ($P=32$, $C=32$) normalized by the number of normal modes $N=2P^{2}$, for different MC sampling times $\tau$.
Top: Patch from (100) FCC plane. 
Bottom: Patch from (111) FCC plane.
Upper panel: $D_\omega{}/N$ scaled by the non-Debye scaling $\omega^3$. 
Lower panel: $D_\omega{}/N$ scaled by the Debye scaling $\omega^2$.
}
\label{fig: DOS scaling evolve}
\end{figure}

We assemble displacement covariance matrices to obtain the Green's function, $\mathcal{G}_{i\alpha{}j\beta}=\mean{u_{i\alpha}u_{j\beta}}/k_{b}T$, for $(111)$ and $(100)$ patches having $P=32$ atoms along each edge.
Here $u_{i\alpha}$ is the displacement of particle `i' along axis $\alpha$ from its equilibrium position and the angle brackets denote averaging over MC sweeps (in one sweep we attempt to move each degree of freedom according to the Metropolis algorithm).
We report results as a function of sampling time `$\tau$', that we define as the number of (post equilibration) MC sweeps per degree of freedom in the patch (there are $2P^{2}$ degrees of freedom in a $P\times{}P$ patch)~\cite{Henkes:2012softMat}.
Note that $\tau$ counts successive MC sweeps and does \emph{not} address the issue of statistical independence between sampled configurations.

We estimate the `decorrelation time' in our MC setup by fitting the energy and  single-particle displacement auto-correlation functions to an exponential function to obtain the characteristic decay time~\cite{NewmanBarkema}.
The decorrelation time is insensitive to size of the FCC crystal $C$. 
We find that the decorrelation time estimated from energy and displacement auto-correlation is roughly 8 and 24 MC sweeps, respectively.
These decorrelation times give some idea about the number of \emph{uncorrelated} samples. 


In figure~\ref{fig: patch MC eigenvalue evolution} we present the inferred frequency spectrum of $\mathcal{G}$ of our $(111)$ patch for various $\tau$ together with the analytical result.
We note that if $\tau<1$ (the number of samples is less than the number of degrees of freedom) we will  necessarily pick up trivial modes (as we see in the spectrum corresponding to $\tau = 0.977$).
Different parts of the frequency spectrum converge at quite different rates: the higher end of the frequency spectrum takes particularly long to converge.
In fact, the change in the inferred spectrum with increasing $\tau$ is so slow that if we only consider data in the top plot of figure~\ref{fig: patch MC eigenvalue evolution} we might be tempted to conclude that the spectrum is converged at about $\tau = 17.58$ since the relative change per unit $\tau$ in the spectrum is quite small.
Indeed, small relative changes in the inferred spectrum are essentially the only tool an experimenter has to declare convergence since an analytical result isn't available~\cite{Ghosh:2010softMat, Kaya:2010science, Henkes:2012softMat}.
The bottom panel of figure~\ref{fig: patch MC eigenvalue evolution} shows the inferred spectrum and the analytical result in the limit of very large $\tau$. 
The data make it clear that convergence to the true spectrum is quite slow.

Figure~\ref{fig: DOS scaling evolve} shows $D_{\omega}$ of for various $\tau$.
Again, we have scaled $D_{\omega}$ by the non-Debye prediction $\omega^{3}$, as well as the Debye law $\omega^{2}$.
The low frequency behavior of the $(100)$ DOS (top) is not severely impacted by the poor statistics and even for relatively small data sets, the $D_\omega /\omega^3$ data shows a much clearer plateau than $D_\omega /\omega^3$.
One would have little case to doubt the predicted non-Debye behavior.
Surprisingly, the $(111)$ patch (bottom plot) shows an \emph{apparent} $\omega^2$ scaling regime at intermediate $\omega$ that tends to disappear only at very large $\tau$.
As we have seen in the previous section, even the true underlying spectrum, at least for these patch sizes, shows very little evidence for any $\omega^3$ scaling regime.
So it appears that the artifacts from the finite statistics act in the same way as the artifacts in the true underlying spectrum to push the result further away from the $\omega^3$ scaling toward an apparent $\omega^2$ scaling.

\begin{figure}[t!]    
\begin{center}
\includegraphics[width=.49\textwidth]{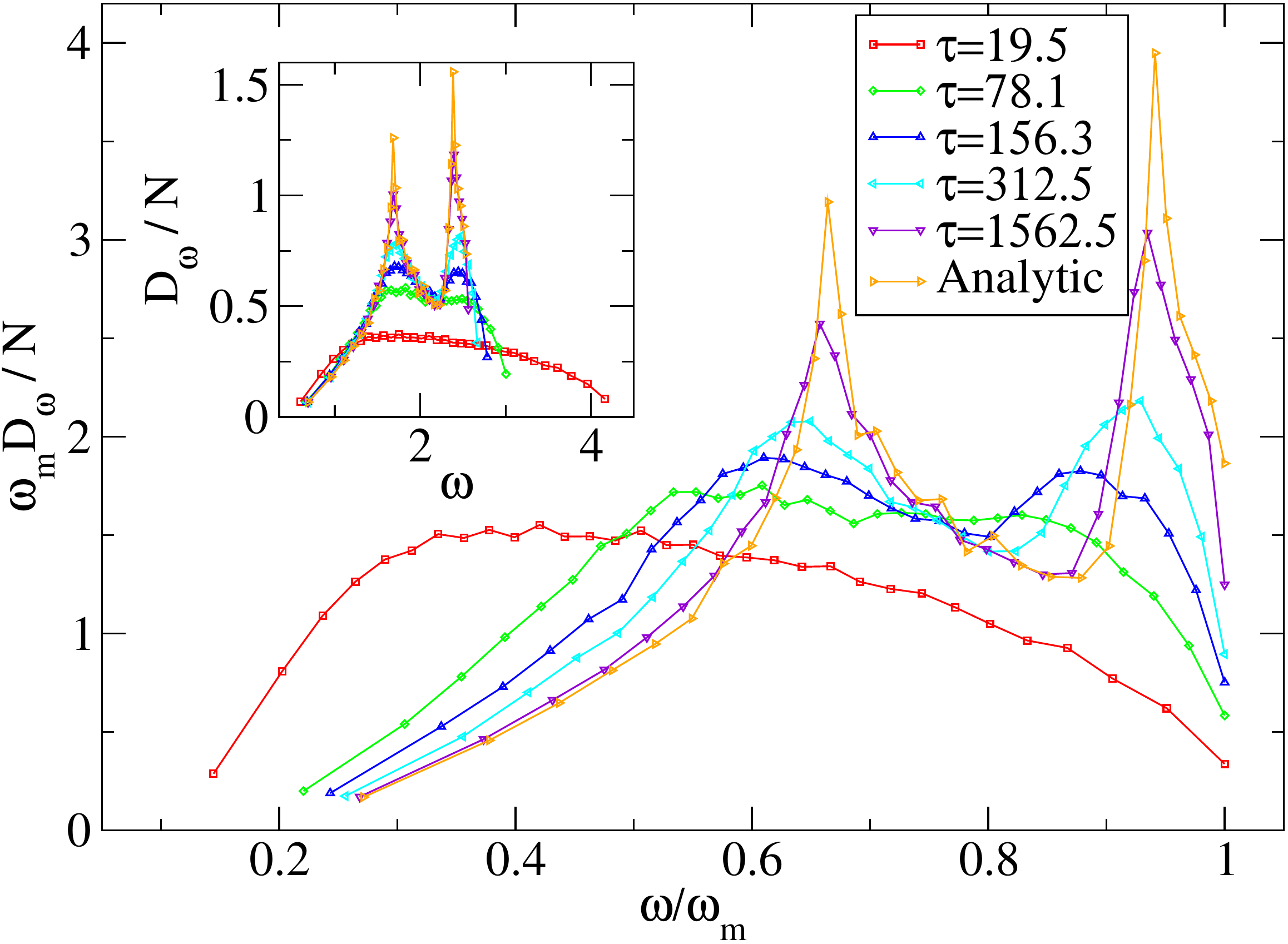}
\end{center}
\caption{
DOS, $D_{\omega}$, of a $(111)$ FCC crystal patches ($P=32$, $C=32$) normalized by $N=2P^{2}$ for different MC sampling times $\tau$.
Inset: $D_{\omega}/N$ as a function of frequency $\omega$.
Main plot: $D_{\omega}/N$ and  $\omega$, non-dimensionalized by the maximum frequency $\omega_{m}$.  
}
\label{fig: DOS evolve}
\end{figure}

To make one final point, we plot the \emph{unscaled} version of the $(111)$ DOS in Figure~\ref{fig: DOS evolve}~\footnote{$(100)$ results are qualitatively the same}.
In the main plot, we normalize frequencies by the maximum observed frequency, $\omega_m$, in the data set as one might do with experimental data, while the inset normalizes the frequencies by the underlying frequency scale in our known Hamiltonian, $\omega_0$.
We see that the inferred $D_{\omega}$, as a whole, shifts to lower frequencies $\emph{relative}$ to the entire spectrum for shorter MC sampling times.
This is essentially due to the fact that $\omega_m$ shifts downward with increasing observation time.
Interestingly, at least with these normalization conventions on the energy scale, the effect of spurious inferred correlations arising from insufficient sampling act in a way very much like true underlying disorder.
For infinitely good statistics, or no underlying disorder, one has sharp van Hove peaks in the DOS.
As the statistical sample is degraded, or one introduces disorder into the underlying model, the van Hove peaks broaden and shift to lower frequencies~\cite{Taraskin:2001prl}.
In effect, one may observe an excess of low frequency modes, or Boson peak, in the inferred spectrum because of insufficient statistics rather than any true underlying disorder. 


\section{Discussion and Summary}
\label{section: discussion}

The ultimate use of the analysis in this paper is to inform experimenters studying glassy or crystalline colloidal suspensions using particle trajectory data obtained from optical microscopy techniques.
As we have shown, there are two issues that much be carefully addressed in analysis of such data: the effects of observing a restricted 2D portion of the system and imperfect statistical information about the displacement correlations.
We first described an analytical approach to understanding projection artifacts for systems governed by a harmonic Hamiltonian and then analyzed MC simulations based on the exact same Hamiltonian to study the statistical artifacts.
The main utility of the approach we described here, and the advantage over previous work, is that both kinds of artifacts -- projection and statistical -- are addressed within a single framework.

As a side benefit, the symmetry properties of the projected Green's function arise as a natural consequence of the symmetries of the embedding FCC lattice and the use of central forces.
We did not need to adjust any free parameters describing the symmetry in the patch, {\it e.g.} as in reference~\cite{Maggs:2011epj-aniso}.
A fully atomistic model also gives a realistic description of modes all the way up to the BZ boundary, whereas other approaches rely on ad-hoc discretization of an elastic continuum. 

We have shown that for any direction in the Brillouin zone, the Green's function of a 2D patch embedded in an FCC crystal has an anomalous non-linear dispersion reation $\omega^2\sim{}q$, inline with earlier, long-wavelength limit, analytical prediction~\cite{Maggs:2012epl, Maggs:2011epj-aniso}. 
Despite the same long wavelength scaling, the dispersion relation is very anisotropic for $(100)$ patches, whereas it is fairly isotropic in the $(111)$ case, at least at long wavelengths. 
Moreover we find that anisotropy  is much more pronounced for the transverse dispersion of $(100)$ FCC patches.
Notably for $(100)$ patches we find that  along certain directions in the BZ, the longitudinal and transverse branches of the dispersion are nearly degenerate all the way to the BZ boundary. 

The non-linear, long wavelength, dispersion relation leads one to expect a non-Debye scaling, $D_\omega\sim \omega^3$ in the DOS of the FCC patches at low frequencies.
Our calculations show the $(100)$ DOS has clear $D_\omega\sim \omega^3$ scaling at low frequencies, however the picture is somewhat ambiguous for $(111)$ patches and does not show a clear $D_\omega\sim \omega^3$ regime.
This is similar to the observations from molecular dynamics simulations of FCC hard sphere systems in reference~\cite{Maggs:2011epj-aniso}.
From the calculated DOS, that we show in figure~\ref{fig: DOS computed}, we see that the deviation from  $\omega^{3}$ scaling at low $\omega$ can arise even with perfect statistical information.  
However, we do expect that this deviation itself is a finite (patch) size effect and that a $D_\omega\sim \omega^3$ scaling regime would emerge for large enough patches.

Next we studied the character of normal modes in the patches.
The DSF gave the contribution of planewaves to normal modes of a given frequency.
In both the $(111)$ and $(100)$ cases, the DSF at a given frequency was essentially determined by equi-$\omega$ contours of the dispersion.  
The strong anisotropy of the $(100)$ dispersion made for broad contributions to the isotropically averaged DSF at a given $\omega$ due to the shape of the dispersion contour.
 
The issue of statistical convergence was more subtle.
We showed that the DOS converges surprisingly slowly with increasingly good statistical estimates of the displacement covariance.
In general, we found that poor statistics smears out the van Hove singularities in the DOS and shifts the peak toward lower frequencies, much in the same way as true disorder.
For the $(111)$ patches, these sampling artifacts can give an apparent low frequency form for the DOS that appears to agree quite well with the conventional Debye result.
However, the DOS becomes closer to the expected non-Debye form with better statistics.  


In the future, we would like to extend our study to harmonic FCC crystals with springs drawn from a distribution (as in refs~\cite{Taraskin:2001prl, Ganter:1998prl}).
It will be important to understand, quantitatively, the competing effects of incomplete statistical information and actual underlying disorder on the inferred spectrum. 
Beyond this, more realistic models for interacting colloidal particles -- such as hard sphere systems -- bring anharmonicity and non-Gaussian behavior.
Understanding all these effects in quantitative ways is necessary before precise connections can be made -- in experiments or simulations --  between the underlying interactions and the observed displacement correlations in small low-dimensional observation windows.


\section*{Appendix}

\begin{appendices}
\section{Fourier space calculation of Green's function}
\label{section: Green's functions analytical expressions}

The Green's function of a crystal $\mathbb{G}=\mathbb{H}^{-1}$, where $\mathbb{H}$ is its Hessian matrix.
For a crystal that has a unit cell with a basis and a pair interatomic potential,  we denote the components of the Hessian matrix by: $H_{i\alpha\mu{}j\beta\nu}$, here $i,j$ label unit cells, $\mu ,\nu$ label atoms in the basis and $\alpha , \beta$ label Cartesian axes.
We consider periodic FCC crystals in a cubic cell with $C$, 4-atom cubic unit cells along each edge. 
This gives us $N=C^3$ unit cells and $M=4N$ atoms in the crystal.   
Translational invariance in the crystal implies that the Hessian is a function of only the \emph{separation}  $\vec{R}_{ij}=\vec{R}_{i}-\vec{R}_{j}$ between unit cells, where $\vec{R}_{i}$ denotes the position of the $i$th.~unit cell, thus giving $H_{i\alpha\mu{}j\beta\nu}=H_{\alpha\mu{}\beta\nu}(\vec{R}_{ij})$
The set of allowed $\vec{R}_{ij}$ is precisely the set of $N$ position vectors, $\vec{R},$ of the unit cells;  suppressing indices labeling pairs of unit cells, we write the components of the Hessian as: $H_{\alpha\mu{}\beta\nu}(\vec{R})$.
This allows us to represent $\mathbb{H}$, which is a $3M\times{}3M$ matrix, in terms of $N$, $12\times{}12$ matrices $\tensor{H}(\vec{R})$, whose components are $H_{\alpha\mu{}\beta\nu}(\vec{R})$.  
Note that having a 4-atom unit cell implies that: $\tensor{H}(\vec{R})=\tensor{H}^{T}(-\vec{R})$, due to translational invariance of the lattice.

Following~\citet{AshcroftMermin}, we take a 3D Fourier transform of the Hessian matrix $\tensor{H}(\vec{R})$ to compute the dynamical matrices according to: \[\fttensor{D}(\vec{k})=\fft{H}.\]
Here $\vec{k}$ is a vector in the cubic reciprocal lattice and the sum runs over the set of $N$ unit cell position vectors $\vec{R}$.
The tilde denotes Fourier transformed operators.
The dynamical matrices $\fttensor{D}(\vec{k})$ are $12\times{}12$ complex, Hermitian matrices.
Due to the identity $\tensor{H}(\vec{R})=\tensor{H}^{T}(-\vec{R})$, we also have: $\fttensor{D}(-\vec{k})=\fttensor{D}^{T}(\vec{k})$.

Next, we numerically invert the $\fttensor{D}(\vec{k})$ to get the Fourier space Green's function $\fttensor{G}(\vec{k})$: \[\fttensor{G}(\vec{k}) = \fttensor{D}^{-1}(\vec{k}).\] 
Since there are 3 trivial translational modes of the crystal, $\fttensor{G}(\vec{0})$ is computed by taking a Moore-Penrose pseudoinverse via a singular value decomposition (SVD).
Finally an inverse Fourier transform of the $\fttensor{G}(\vec{k})$ gives us the complete real space Green's function $\tensor{G}(\vec{R})$: \[\tensor{G}(\vec{R}) = \ifft{G}.\]
In our implementation, we perform the forward Fourier transform as an explicit sum over $\vec{k}$ as very few of the $\tensor{H}(\vec{R})$ are non-zero due to short range interactions; e.g.~in the case of only nearest neighbor interactions there are at most $3^3=9$ non-zero $\tensor{H}(\vec{R})$. 
However, the inverse Fourier transform is performed via a fast Fourier transform routine.

\end{appendices}
\vspace{2mm}
\acknowledgments
We thank Michael Widom for useful discussions at various stages during this work.
This material is based upon work supported by the National Science Foundation under Award Number NSF-DMR-1056564.

\bibliography{ColloidRefs}
\bibliographystyle{apsrev}
\end{document}